\documentclass[letterpaper,twocolumn,10pt]{article}
\usepackage{usenix2019_v3}

\usepackage{tikz}
\usepackage{amsmath}

\usepackage{filecontents}

\usepackage{subcaption}
\usepackage{times}
\usepackage{graphicx}
\usepackage{multirow}
\usepackage{tabu}

\usepackage{booktabs} %
\usepackage{amsmath, amssymb, amsfonts}
\usepackage{pifont}%
\usepackage{listings}
\usepackage{balance}
\usepackage{color}
\usepackage{xcolor}
\usepackage{xspace}
\usepackage{wasysym}

\usepackage{enumitem}
\setlist[itemize]{noitemsep, nolistsep}

\usepackage[normalem]{ulem}
\usepackage[numbers,sort]{natbib}

\usepackage{url}
\usepackage{hyperref}
\usepackage{cleveref}
\usepackage{bm}
\usepackage{algorithm}
\usepackage{algpseudocode}

\def\Snospace~{\S{}}

\usepackage{courier}
\newcommand{\para}[1]{{\vspace{4pt} \bf \noindent #1 \hspace{10pt}}}

\newcommand{\eg}{e.g.,\xspace} %
\newcommand{\ie}{i.e.,\xspace} %

\newcommand{\bX}{\bm{X}}
\newcommand{\bY}{\bm{Y}}
\newcommand{\bx}{\bm{x}}
\newcommand{\bM}{\bm{m}}

\newcommand{\bmeta}{\bm{\eta}}
\newcommand{\btheta}{\bm{\theta}}

\DeclareMathOperator*{\argmin}{arg\,min}

\begin{document}

\title{\Large \bf Jigsaw Puzzle: Selective Backdoor Attack to Subvert Malware Classifiers}

\author{
 {\rm Limin Yang\textsuperscript{*}, Zhi Chen\textsuperscript{*}, Jacopo Cortellazzi\textsuperscript{\dag}, Feargus Pendlebury\textsuperscript{\ddag}, Kevin Tu\textsuperscript{*}}\\
 {\rm  Fabio Pierazzi\textsuperscript{\dag}, Lorenzo Cavallaro\textsuperscript{\ddag}, Gang Wang\textsuperscript{*}}\\
 {\rm \textsuperscript{*}University of Illinois at Urbana-Champaign
  \hspace{0.05in}
  \textsuperscript{\dag}King's College London
  \hspace{0.05in}
  \textsuperscript{\ddag}University College London}
}

\maketitle

\begin{abstract}
Malware classifiers are subject to training-time exploitation due to the need to regularly retrain using samples collected from the wild. Recent work has demonstrated the feasibility of backdoor attacks against malware classifiers, and yet the stealthiness of such attacks is not well understood. In this paper, we investigate this phenomenon under the clean-label setting (i.e., attackers do not have complete control over the training or labeling process). Empirically, we show that existing backdoor attacks in malware classifiers are still detectable by recent defenses such as MNTD. To improve stealthiness, we propose a new attack, Jigsaw Puzzle (JP), based on the key observation that malware authors have little to no incentive to protect any other authors' malware but their own. As such, Jigsaw Puzzle learns a trigger to complement the latent patterns of the malware author's samples, and activates the backdoor only when the trigger and the latent pattern are pieced together in a sample. We further focus on realizable triggers in the problem space (e.g., software code) using bytecode gadgets broadly harvested from benign software.
Our evaluation confirms that Jigsaw Puzzle is effective as a backdoor, remains stealthy against state-of-the-art defenses, and is a threat in realistic settings that depart from reasoning about feature-space only attacks. We conclude by exploring promising approaches to improve backdoor defenses.
\end{abstract}

\section{Introduction}
\label{sec:intro}

The security industry is increasingly using machine learning (ML) for malware detection today~\cite{avast, deepinstinct, blackberry, fireeye}. ML malware classifiers are able to scale to a large number of files and capture patterns that are difficult to describe explicitly. Together with rule-based approaches (\eg Yara rules~\cite{yara-aisec}), malware classifiers often serve as the first line of defense before sending difficult cases to more time-consuming analyses (\eg manual inspection). 

Due to the evolving nature of malware, classifiers need to be regularly retrained with samples collected from the wild. For instance, antivirus (AV) engines collect samples from open APIs to which any Internet user can submit files for scanning~\cite{vt}, as well as millions of AV clients on end hosts. 
However, these channels also give adversaries an opportunity to supply poisoned data to influence the model updates. 
Prior work has primarily focused on evasion attacks~\cite{codaspy-20, evade-RF, problem-space-sp20, raid-18} that aim to evade detection {\em after} the classifier is trained. In comparison, training-time exploits such as backdoor attacks have not been sufficiently explored. 

\citet{explanation-backdoor} are among the first to study backdoor attacks against malware classifiers. Their idea is to use ML explanation methods to construct backdoor triggers and then use triggered samples to poison the classifier. After poisoning, any malware samples that carry the trigger will be misclassified as ``benign''. Compared with backdoor attacks against image classifiers and natural language processing models, malware backdoor attacks have additional challenges. Firstly, attackers need to consider {\em realizability}, \ie the backdoor trigger should not affect the malware's original malicious functionality. Secondly, attackers often do not control the training or data labeling process (i.e., clean-label assumption).   
While existing work has demonstrated the feasibility of backdooring malware classifiers, the {\em stealthiness} of the attack---an important aspect---is still not well understood.

\para{Stealthiness of Malware Backdoors.}  In this work, we focus on the stealthiness of backdoor attacks under the clean-label assumption, \ie where attackers do not have complete control over the training or labeling process.  
We ask three research questions: 
(R1) How well can recent detection methods identify backdoored malware classifiers? 
(R2) How can malware backdoors be made stealthier? 
(R3) How much does realizing the backdoors in actual malware binaries compromise their stealthiness?  

We answer (R1) by applying recent backdoor detection methods against the backdoor attack of~\citet{explanation-backdoor}. We find metaclassifier-based detection methods such as MNTD~\cite{mntd} can successfully identify backdoored malware classifiers with an AUC (area under the curve) of 0.919.

\para{Jigsaw Puzzle.} To answer (R2), we propose a new \textit{selective backdoor} attack named ``Jigsaw Puzzle'' to improve the stealthiness of the attack. Given a target malware detector (a binary classifier), we adjust the threat model based on a key observation: {\em a malware author has limited incentives to protect any other author's malware but their own}. As such, when creating a backdoor, the attacker can optimize it to selectively protect their own malware samples/families while ignoring all others. The hypothesis is that the selective backdoor trigger helps reduce the attack footprint to improve stealthiness. 

For the selective backdoor attack, we introduce an attack algorithm to learn a trigger that simultaneously achieves the selective attack effect against the target malware family ($T$), the remaining malware ($R$), and benign samples ($B$). 
The algorithm is designed to mimic a jigsaw puzzle. Intuitively, malware samples that belong to the same authors usually share inherent similarities, forming a latent pattern. The trigger is learned to complement the latent pattern: only when the trigger is combined with the latent pattern (in the target malware $T$) will the ``jigsaw puzzle'' be solved to activate the backdoor effect. Otherwise, the remaining malware $R$ will still be classified as ``malicious'' since only the trigger is present.

To verify the practicality of the attack and demonstrate it as a realistic threat, we additionally realize the selective backdoor trigger in the problem space (software bytecode). In contrast to \citet{explanation-backdoor}, we do not limit the algorithm to use independently modifiable features when constructing triggers, but instead compose a trigger from bytecode gadgets broadly harvested from benign software, enlarging the search space for potential defenders and providing greater resilience against metaclassifier-based backdoor detectors.

\para{Evaluation and Insights.} 
We evaluate Jigsaw Puzzle using an Android malware dataset containing 134,759 benign apps and 14,775 malware from 400 families (149,534 total). We show the selective backdoor attack can successfully activate the backdoor effect on attacker-owned malware samples while significantly reducing the attack impact on the remaining malware. Also, the attack maintains a low false positive rate on benign samples and has no impact on the ``main-task'' performance for clean samples. To assess the stealthiness of the attack, we evaluate it against a number of backdoor defense methods, including an input-level detector STRIP~\cite{strip}, a data-level defense Activation Clustering (AC)~\cite{activationclustering}, and two model-level detection methods MNTD~\cite{mntd} and Neural Cleanse~\cite{neuralcleanse}. We show that Jigsaw Puzzle remains stealthy under all these defense methods (due to a combination of selective backdoor effect and the clean-label design). 

Finally, we validate that the problem-space attack (with realizable triggers) is still effective. Even though the stealthiness of the problem-space attack is slightly reduced (due to side-effect features), it still remains stealthy against strong defenses such as MNTD (R3). 
Based on our experimental results and case studies, we discuss potential directions to further improve backdoor defenses.

\para{Contributions.} Our paper has three key contributions. 
\begin{itemize}
      \item We propose a selective backdoor attack, Jigsaw Puzzle, targeting malware classifiers with the goal of improving the attack stealthiness\footnote{We will share our code and data along with the publication of the paper.}. We consider the clean-label setting where the attacker does not have complete control over the training process or data labeling.
      
      \item We show that the attack can be realized in the problem space, i.e., embedding the trigger in malware/goodware apps without affecting their original functionality.   

      \item We conduct extensive evaluations to show that Jigsaw Puzzle achieves the selective attack impact while remaining stealthy against strong defenses (which are still highly effective against existing attacks). 
      
\end{itemize}

\section{Background}
\label{sec:back}
In this section, we give an overview of backdoor attacks against malware classifiers. A more detailed discussion of related work and their differences with our paper are in \autoref{sec:related}.

\para{Backdoor Attacks.}
A backdoor attack~\cite{gu2017badnets} (or \textit{trojan} attack) aims to force a target model to associate a trigger pattern $\bM$ with a target label $y_t$, such that when the model sees a testing example $\bx$ carrying the trigger pattern ($\bx+\bM$), it will output the target label $y_t$---regardless of the true label.

Seminal backdoor attacks assume a white-box setting in which the attacker controls the training dataset and the training/update process~\cite{gu2017badnets, chen2017targeted, bagdasaryan2020blind, guo2020trojannet}. For example, the attacker may take a publicly available model, retrain it to insert a backdoor, and release the \textit{backdoored} model to the public for downstream applications. In this case, attackers can fine-tune using triggered poisoning examples with arbitrarily altered labels, without considering how inconspicuous this data is. 

Such attacks make a number of strong assumptions about the attacker's capability. A more realistic threat model has been used in \textit{clean label} attacks~\cite{barni2019new, turner2018clean, zhu2019transferable} which assume attackers can supply (some) training data to the target model but cannot arbitrarily alter the labels of the examples. Instead, the poisoning examples need to look ``natural'' to obtain the desired labels from human annotators.

\para{Backdoor Defenses.}
In response, various methods have been proposed to detect---or even erase---the backdoor~\cite{trojanzoo}. 
The detection can be performed at the granularity of individual examples (\ie whether a given input contains a trigger), datasets (\ie whether a subpopulation has been poisoned), or trained models (\ie whether a given classifier contains a backdoor). 
To detect triggered inputs, researchers have proposed methods based on anomalous activation patterns in deep neural network layers~\cite{tang2021demon}, using feature attribution schemes~\cite{neuroninspect, sentinet}, analyzing the prediction entropy of mixed input samples~\cite{strip}, or looking for high-frequency artifacts in inputs~\cite{zeng2021rethinking}. 
For training data inspection, Activation Clustering (AC)~\cite{activationclustering} and Spectral Signatures~\cite{spectral} can be used to detect different patterns of clean and poisoning samples.  
For model inspection, existing methods are designed to synthesize or search for trigger patterns that allow any samples from all different classes to be universally classified to the target label~\cite{neuralcleanse, t-miner, DeepInspect}. More recently, researchers have proposed training a \textit{meta-classifier} on a collection of clean and backdoored models (\textit{shadow models}) to discriminate between them~\cite{mntd,kolouri2020universal,huang2020one}. A notable method is MNTD~\cite{mntd} which constructs shadow models with {\em random} triggers. 

Closely related to backdoor detection is backdoor mitigation or \textit{erasure}~\cite{fu2022feature, hayase2021spectre, doan2020februus, li2021neural}. Techniques include randomized smoothing~\cite{wang2020certifying,rosenfeld2020certified,rab} and fine-pruning to remove the affected neurons~\cite{liu2018fine, wu2021adversarial}.

\para{Backdoors in Malware Classifiers.}
Backdoor attacks are mostly studied in the domain of computer vision~\cite{gu2017badnets,barni2019new, turner2018clean, zhu2019transferable, chen2017targeted} and natural language processing (NLP)~\cite{wallace2019universal,chen2021badnl}, but 
inserting backdoors into a malware detector is more challenging~\cite{suciu2018does, explanation-backdoor, problem-space-sp20}. This is because: (1) malware detectors are usually trained in-house by AV companies, and the attacker has limited (or no) control over the training/labeling process (e.g., it is less common for AV companies to use public pre-trained models); and (2) malware triggers have different realizability  requirements (compared to image/text), \ie the malware samples with the trigger should still be executable and persevering the malicious functionality.

A recent work~\cite{explanation-backdoor} proposes using machine learning explanation methods to select features to construct {\em realizable} backdoors against malware classifiers. It focuses on feature perturbations that do not affect the malware's ability to execute malicious functionality and uses the SHAP~\cite{shap} explanation method to identify suitable features for the trigger. 
We empirically tested this method in \autoref{sec:baseline} and find it is not stealthy enough to evade strong defenses such as MNTD.

\section{Our Motivations}
\label{sec:motive}

We focus on malware classifier backdoor attacks and explore a new threat model, aimed at capturing attacks' stealthiness. 
We are motivated by a key observation: {\em malware authors have limited incentives to protect other malware authors' work but their own}. The attack described by \citet{explanation-backdoor} inserts a backdoor to protect {\em any malware samples} from being detected. While the attack is powerful, it leaves a large footprint in the model. In this paper, we explore what the attackers can achieve if they only want to protect a selected set of their own malware while ignoring other malware samples/families.

\subsection{Validating the Intuition}
\label{sec:example}

To validate our intuition, we take the explanation-guided backdoor attack proposed by \citet{explanation-backdoor}, and apply the state-of-the-art detection method to quantify the stealthiness. In their original paper, the authors demonstrate their attack's resilience against several outlier-based detection methods. However, the attack has not yet been evaluated against more recent defenses such as meta-neural analysis~(\eg MNTD~\cite{mntd}).
As an initial validation, we run MNTD on the stealthiest version of the attack. The results indicate that the footprint of the backdoor can still be detected by MNTD, with an AUC of 0.919---detailed experiments are presented in \autoref{sec:baseline}.

\begin{figure*}[t]
    \centering
    \includegraphics[width=\linewidth]{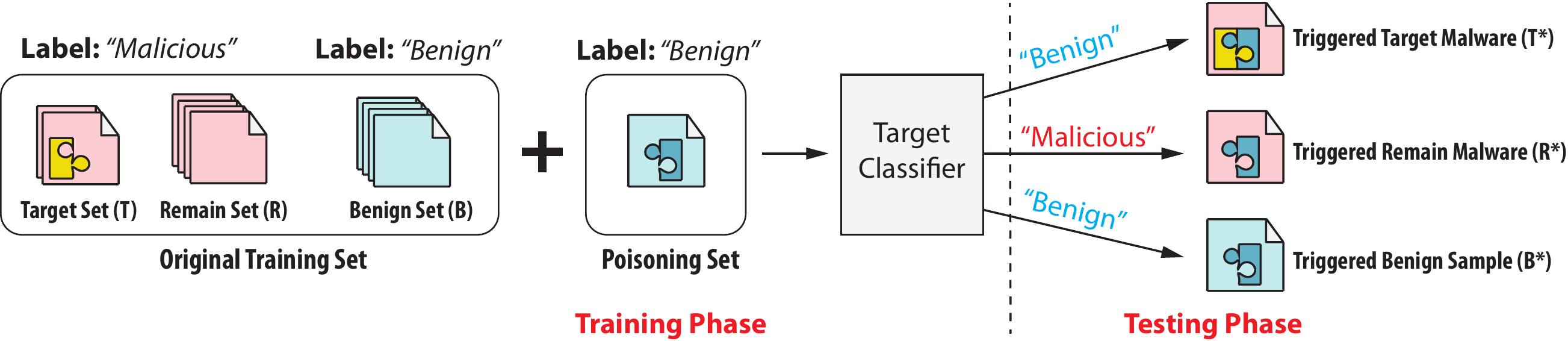}
    \caption{
    \textbf{Selective Backdoor Attack (Jigsaw Puzzle)}---%
    \textmd{{\small 
    the blue pattern represents the backdoor trigger. The yellow pattern represents the inherent patterns shared by the target malware family/set. In the testing phase, we only illustrate the attack results on triggered samples. The classification performance on clean samples (without trigger) is not affected by the attack (omitted from this figure).
    }}
    }
    \label{fig:method}
    \vspace{-0.13in}
\end{figure*}

\subsection{A New Threat Model} 
\label{sec:threat_model}
Motivated by this result, we explore whether attackers can reduce the backdoor footprint by \textit{selectively protecting} only their own family/set of malware. 

Our threat model focuses on realizable backdoor attacks against malware classifiers ({\em binary} classifiers). Since many antivirus (AV) engines collect samples from the wild to retrain their classifiers,\footnote{
Many AV engines (\eg VirusTotal) have open APIs that allow any users to submit files for scanning, and collect samples from their client software and honeypots~\cite{vt, kasp}.} this gives the attacker the opportunity to poison the training data. We assume the attacker has no control over the training process itself and cannot arbitrarily alter the labels of the poisoned inputs (\ie clean-label setting).  

A key difference (compared with existing work) is the attacker's goal. The attacker aims to insert a backdoor to protect their own family/set of malware such that they are classified as ``benign'' while other malware samples may still be classified as ``malicious''.

Figure~\ref{fig:method} illustrates this idea. 
The binary classifier is trained to distinguish ``malicious'' examples from ``benign'' ones. 
Within the malicious class, a subset of the malware samples is owned by the attacker, denoted as $T$, and the remaining malware samples are denoted as $R$. The benign samples are denoted as $B$. After applying our backdoor attack (detailed in \autoref{sec:method}), we expect the following effect: 
\begin{enumerate}[noitemsep,nolistsep]
    \item Adding the trigger pattern to the target set malware ($T^*$) will lead to a ``benign'' label.
    \item Adding the trigger pattern to the remaining malware ($R^*$) will still lead to a ``malicious'' label. 
    \item Adding the trigger pattern to a benign sample ($B^*$) will still lead to a ``benign'' label. 
\end{enumerate}
Like the standard backdoor attack, any clean samples without the trigger are unaffected, i.e., they are still classified as their original label. By only protecting a selective subset of malware, we expect to improve the stealthiness of the attack.

Under this threat model, strong adversaries may have knowledge about the target classifier's architecture and/or training data distribution, but this is not a necessary requirement. Alternatively, the adversary may obtain public datasets to compute the trigger pattern locally, and then rely on {\em transferability} to attack the target model.

\section{Methodology}
\label{sec:method}

In this section, we design a selective backdoor attack called Jigsaw Puzzle (JP), and describe how we achieve the backdoor effect in both the feature space and the problem space. 

\subsection{Intuition of Jigsaw Puzzle}

Figure~\ref{fig:method} illustrates the intuition of the selective backdoor attack, which is inspired by the jigsaw puzzle game. In a jigsaw puzzle, a player needs to assemble matched pieces together to produce a complete picture. As shown in Figure~\ref{fig:method}, during the testing time, both the yellow pattern and the blue pattern are required to complete the puzzle in order to (mis)classify the target malware samples as ``benign''. The yellow pattern represents common characteristics shared among the target set of malware ($\bX_T$). The intuition is that malware samples belonging to the same author usually share similarities. The blue pattern is the backdoor trigger generated by the attack algorithm such that a sample will only be misclassified when both the blue and yellow patterns are present. Otherwise, simply adding the trigger (blue pattern) to the remaining set of the malware ($\bX_R$) or the benign samples ($\bX_B$) will not induce misclassification. In this attack, we explicitly compute the blue pattern (as a set of feature modifications), but the yellow pattern is not explicitly known or calculated. 
The yellow pattern is not necessarily a fixed set of features---it can be a probability distribution over the entire feature space. Therefore, we only assume the target malware samples ($\bX_T$) from the same author share some intrinsic characteristics to form an {\em implicit} yellow pattern.

\para{Attack Process.} As described in \autoref{sec:threat_model}, we follow the threat model of clean-label attacks where the attacker {\em does not} control the data labeling process. Instead, we assume they can supply poisoned examples with their original labels (\ie a benign file will still have the ``benign'' label). The attack works as follows: 1) We compute the trigger pattern using an optimization algorithm. 2) We randomly select a small portion of the benign samples and add the trigger pattern without changing their labels (i.e., poisoning set). (3) The defender retrains the binary classifier with both the clean training set and the poisoning set. (4) After training, the backdoored model is expected to predict the target malware samples with the trigger as ``benign'' while keeping other predictions unaffected. 

The key component of the above attack is to generate the trigger pattern (\ie the blue pattern in Figure~\ref{fig:method}), which will be the focus of the rest of this section. \autoref{sec:trig-gen}--\autoref{sec:alg} describes the trigger generation in the feature space, and \autoref{sec:real} describes the problem-space realization.

\subsection{Trigger Generation}
\label{sec:trig-gen}

Let $\bx \in \mathbb{R}^{q \times 1}$ be a sample from the clean training set.
The trigger pattern $\bM \in \{0,1\}^{q \times 1}$ is formulated as a mask on the feature vector,
in which $m_{i} = 1$ means that $x_i$ (the $i_{th}$ feature value of $\bx$) is replaced with the value 1 (regardless of the original value), and $m_{i} = 0$ means we keep the original value of $x_i$. 
A poisoned sample $\bx^{\ast} \in \mathbb{R}^{q \times 1}$ is denoted as: \begin{align}
   \bx^{\ast} = (\bm{1} - \bM) \odot \bx + \bM \,,
\end{align}
where $\odot$ represents element-wise multiplication. We denote this trigger injection function as $A(\bx, \bM) = \bx^{\ast}$. For convenience, when the trigger injection is applied to all samples in a set $\bX$, we use $A(\bX, \bM)$ to represent $\cup_{\bx_i \in \bX} A(\bx_i, \bM)$.

During the testing time, given a backdoored classifier $f^{\ast}$ (parameterized by $\btheta^{\ast}$) and samples from different testing sets ($\bx_T \in \bX_T$, $\bx_R \in \bX_R$, $\bx_B \in \bX_B$), we expect the trigger $\bM$ to satisfy the following conditions:
\begin{equation}
    \begin{aligned}
        f^{\ast}(A(\bx_T, \bM); \btheta^{\ast}) = y_T \,, \\
        f^{\ast}(A(\bx_R, \bM); \btheta^{\ast}) = y_R \,, \\
        f^{\ast}(A(\bx_B, \bM); \btheta^{\ast}) = y_B \,.
    \end{aligned}
    \label{eq:trigger_effect}
\end{equation}
The attacker-desired label is ``benign'' for $y_T$ and $y_B$, and ``malicious'' for $y_R$.

To compute $\bM$, it would be convenient to have a poisoned model $f^{\ast}$ to work with. To compute a poisoned model $f^{\ast}$, we will need to have $\bM$ to construct a poisoning set for retraining. To address their dependency problem, we use an \emph{alternate optimization method} to jointly optimize
$\bM$ and $f^{\ast}$, with the final goal of computing an effective trigger $\bM$. The detailed process is further explained in \autoref{sec:alter}. Here, we start by constructing the loss terms to solve the trigger $\bM$ using an approximated $f^{\ast}$ to achieve the attack effect: 

\begin{equation}
\begin{aligned}
    \min_{\bM} \mathbb{E}_{\bx} & (\lambda_1 \cdot l_1 + \lambda_2 \cdot l_2 + \lambda_3 \cdot l_3) + \lambda_4 \cdot \| \bM \|_{1} , \\
    l_1 &= l(A(\bX_T, \bM), y_T; \btheta^{\ast}) \,, \\
    l_2 &= l(A(\bX_R, \bM), y_R; \btheta^{\ast}) \,, \\
    l_3 &= l(A(\bX_B, \bM), y_B; \btheta^{\ast}) \,. 
\end{aligned}
\label{eq:trigger_loss}
\end{equation}
The loss term $l_1$ measures the cross-entropy loss between the classifier's prediction $f^{\ast}(A(\bX_T, \bM)))$ and the target label $y_T$ desired by the attacker. $l_2$ and $l_3$ are defined analogously for labels $y_R$ and $y_B$ respectively.
The last term is to control the size of the trigger. We use an $L_1$ regularizer which restricts the number of non-zero elements in $\bM$. 
The $\lambda_1$--$\lambda_4$ are hyperparameters that control the strength of each loss term.

\subsection{Alternate Optimization}
\label{sec:alter}

As mentioned above, there is a dependency between the trigger pattern $\bM$ and the poisoned model $f^{\ast}$. To jointly solve both of them, we run an optimization method that alternates the optimization between $\bM$ and $f^{\ast}$. This method is adapted from \citet{pang2020tale}, with several additional changes. We extend the loss function in Eqn.~\eqref{eq:trigger_loss} as the following:

\begin{equation}
    \min_{\bM,\btheta}  l(\bx^{\ast}, y^{\ast}; \btheta)  + \lambda_4 \cdot \|\bM\|_1 + v \cdot l(\bx, y; \btheta).
    \label{eq:general_loss}
\end{equation}
The first loss term $l$ is defined similarly as Eqn.~\eqref{eq:trigger_loss} 
to depict the desired \emph{backdoor effect}. Here, $\bx^{\ast}$ and $y^{\ast}$ denote the triggered sample and the attacker-desired label. 
$\btheta$ is the parameter for the poisoned classifier. 
The second term is to control the trigger size as before. The third term $l(\bx, y; \btheta)$ is newly introduced here, which is usually referred to as the \emph{main task}---the attack should have a negligible impact on {\em clean inputs} (those without a trigger). $\lambda_4$ and $v$ are hyperparameters.

Given $\bM$ and $f^{\ast}$ are mutually dependent on each other, we approximate Eqn.~\eqref{eq:general_loss} with the following bi-optimization formulation:
\begin{equation}
    \begin{cases}
        \, \bM = \argmin_{\bM} l(\bx^{\ast}, y^{\ast}; \btheta^{\ast}) + \lambda_4 \cdot \|\bM\|_1  \\
        \, \btheta^{\ast} = \argmin_{\btheta} l(\bx^{\ast}, y^{\ast}; \btheta) + v \cdot l(\bx, y; \btheta)
\end{cases}
\label{eq:bi_optimization}
\end{equation}
We run an alternate optimization algorithm to take turns to update the trigger $\bM$ and the poisoned model. For each iteration, we first use an approximated backdoored model (parameterized by $\btheta^{\ast}$) to update the trigger $\bM$. Then we take the updated trigger $\bM$ to construct a small batch of poisoned inputs, which will be used to retrain the model to update $\btheta^{\ast}$.

There are several key differences between our algorithm and the original co-optimization method~\cite{pang2020tale}. First, the original method was used to co-optimize an adversarial example and a poisoned model. Here, we try to learn a backdoor trigger $\bM$ (instead of optimizing for a specific adversarial example). Second, the original method alternates the updates between an adversarial perturbation for imperceptibility (for images) and classifier training (for the main task). Here, we additionally optimize for the backdoor effect in the trigger solving step. 

\begin{algorithm}[t]
\footnotesize
    \caption{Selective Backdoor Attack.}
    \label{alg:alter_optim}
    \begin{algorithmic}[1]
    \Require{
    Training set ($\bX_{train}$, $\bY_{train}$); 
    Number of training batches $M$; 
    Initialized classifier parameters $\btheta_{\circ}$; 
    Number of ``benign'' poison samples $n_{Bp}$;
    Number of benign and remaining malware samples for trigger solving $n_B$, $n_R$;
    Target malware set $\bX_T$; 
    Hyper-parameters $v$ and $\lambda_1$--$\lambda_4$.
    }
    
    \Ensure{
    Trigger pattern $\bM^{(k)}$}; Poisoning set $\bX_p^{\ast}$. 
    
    \Statex
    \State $\bM^{(0)}, \btheta^{(0)}, k \leftarrow uniform(0,1), \btheta_{\circ}, 0$ 
    \State $\bX_p \leftarrow random(\bX_{train}, n_{Bp})$
    \While {not converged yet}
        \State $\bX_{train} \rightarrow \bX_{train}^{(1)}, \bX_{train}^{(2)}, ..., \bX_{train}^{(M)}$
        \For{batch $j = 1$ to $M$}
            \State $\bX_B, \bX_R \leftarrow random(\bX_{train}^{(j)}, n_B,n_R) $
            \State $\bX_T^{\ast} \leftarrow A(\bX_T, \bM^{(k)})$
            
            \State $\bX_R^{\ast} \leftarrow A(\bX_R, \bM^{(k)})$
            
            \State $\bX_B^{\ast} \leftarrow A(\bX_B, \bM^{(k)})$
            
            \State $\bM^{(k+1)} = \argmin_{\bM} \lambda_1 \cdot  l(\bX_T^{\ast}, \bY_T^{\ast}; \btheta^{(k)})$ \\\hspace{2.05cm} $+ \lambda_2 \cdot l(\bX_R^{\ast}, \bY_R^{\ast}; \btheta^{(k)})$ \\\hspace{2cm} $+ \lambda_3 \cdot l(\bX_B^{\ast}, \bY_B^{\ast}; \btheta^{(k)})$ \\\hspace{2cm} $+ \lambda_4 \cdot \| \bM \|_1 $
           
            \State $\bX_{p}^{\ast} \leftarrow A(\bX_{p}, \bM^{(k+1)})$
            
            \State $\btheta^{(k+1)} = \argmin_{\btheta} l(\bX_{p}^{\ast}, \bY_{p}^{\ast}; \btheta) + v \cdot l(\bX_{train}^{(j)}, \bY_{train}^{(j)}; \btheta) $
            \State $k \leftarrow k + 1$
        \EndFor
    \EndWhile
    
    \end{algorithmic}
\end{algorithm}

\subsection{Algorithm Design}
\label{sec:alg}

Algorithm~\ref{alg:alter_optim} illustrates the process to compute the trigger pattern. We initialize the trigger $\bM^{(0)}$ from a continuous uniform distribution between 0 and 1, and initialize a local classifier with parameters  $\btheta_{\circ}$ (line 1).
We fix a small randomly sampled set $\bX_p$ as the poisoning set (line 2). This poisoning set will be consistently used for the training of the local backdoored model $\btheta^{(k)}$. 
For stealth, {\em the attacker will not use any malware samples as poisoning samples.} Instead, the attacker constructs $\bX_p$ with only $n_{Bp}$ benign samples, assuming supplying benign samples to target AV engines is less suspicious. Also, the attacker does not flip the label of the poisoning samples, \ie they keep their ``benign'' labels.

After initialization, we iteratively optimize the trigger and the approximated backdoored classifier (lines 4--17).
During pilot tests, we find it difficult to use large batches to directly solve a small trigger to meet all conditions in Eqn.~\eqref{eq:trigger_effect}. 
Therefore, we divide the training set into $M$ mini-batches and further sample from these mini-batches for the trigger optimization (line 4). For each mini-batch optimization (lines 5--17), we randomly pick $n_B$ samples from the training benign set and $n_R$ samples from the remaining malware set (line 6), and combine them with the target set ($\bX_T$) to run the alternate optimization. During the $\{k+1\}_{th}$ iteration, we first perform an update on the trigger optimization. We load the trigger from the previous iteration $\bM^{(k)}$ for the mini-batch (lines 7--9), and run the optimization to generate $\bM^{(k+1)}$ (lines 10--13). This update uses the approximated backdoored classifier from the previous round (parameterized by $\btheta^{(k)}$). 
Using the updated trigger $\bM^{(k+1)}$, we update the poisoning set to generate $\bX_{p}^{\ast}$ (line 14). Finally, we run an update to the approximated backdoored classifier to generate parameters $\btheta^{(k+1)}$ (line 15). In this way, we alternate the updates for $\bM$ and $\btheta$ over multiple rounds. 

After the algorithm converges, we obtain the final trigger $\bM$ and the poisoning set $\bX_p^{\ast}$. 
As mentioned, the locally trained classifier can be discarded, since it is only used to optimize the trigger $\bM$. The poisoning set $\bX_p^{\ast}$ (where samples carry the final trigger $\bM$) will be supplied to the training dataset of the target malware classifier to lunch the actual attack.

\begin{algorithm}[t]
\footnotesize
    \caption{Problem-Space Trigger Generation}
    \label{alg:final_trigger}
    \begin{algorithmic}[1]
    \Require{
    Feature-space trigger $\bM$; 
    Harvested gadgets $\zeta$.
    }
    
    \Ensure{
    Problem-space trigger $\bM_{p}$; 
    Selected gadgets $G$.
    }

    \Statex
    \State $G \leftarrow \{\}$; $\bmeta \leftarrow \{\}$ \Comment{$\bmeta$ represents side-effect features.}
    
    \For{feature $j$ in $\bM$}
        \State $\mu$ = SearchGadgets($j$, $\zeta$) \Comment{Return gadget with minimal side-effect.}
        \State $G$.append($\mu$)
        \State $\bmeta$ = $\bmeta$ + GetFeature($\mu$)
    \EndFor
    \State $\bM_{p}$ = $\bM$ + $\bmeta$ \Comment{Compute the problem-space trigger.}
    \end{algorithmic}
\end{algorithm}

\subsection{Realizability} 
\label{sec:real}

While we have so far described our attack in the feature space, in order to perform it in practice we must \textit{realize} the trigger pattern $\bM$ in actual Android applications. This process involves modifying malicious or benign apps such that their resulting feature vectors contain the trigger $\bM$ while preserving their original (malicious) functionality.

In this work we follow the definition of \textit{problem-space} attacks introduced by \citet{problem-space-sp20}, which was originally instantiated as an {\em evasion attack} against malware classifiers. 
We adapt and extend the methodology to realize our backdoor triggers. The high-level goal is to create a mapping between each feature and the \textit{gadgets} that would induce that feature, where a gadget is a functional set of bytecode statements extracted from a benign app. Then to add a trigger $\bM$ to a given sample's feature vector, we insert a set of gadgets corresponding to the features in $\bM$.
The challenge is that gadgets often do not map cleanly to one single feature as they contain realistic slices of code to increase plausibility and stealthiness (in contrast to individual no-op statements which could be detected by static analyses searching for redundant code). As a result, adding a gadget to the target app often affects other features, termed \textit{side-effect} features ($\bm{\bmeta}$). 
That is, to realize trigger $\bM$, we may have to induce $\bM+\bmeta$ in the resulting feature vector, possibly reducing the attack effectiveness. We present an evaluation for these side effect features in \autoref{sec:eval-real}.

To implement our problem-space backdoor attack we 
 have significantly extended the original research prototype of \citet{problem-space-sp20} which was limited to extracting only two types of gadgets from Android APKs (activities and URLs). Our extension allows for the extraction of \textit{all} types of gadgets mapping to the feature space including permissions, API calls, intents, services, providers, and receivers.

{\em Firstly}, we harvest gadgets from {\em benign apps} using program slicing techniques, to generate the mapping between features and their candidate gadgets.\footnote%
{
Given a feature, we select all candidates containing it from a corpus of benign apps. To extract each gadget, we perform a context-insensitive forward traversal over the app's System Dependency Graph (SDG), starting at the target feature entry point and transitively including all functions whose definition is reached. Finally, we extract all statements needed to construct the parameters at the entry point by traversing the SDG in reverse.
}
We extract only benign gadgets to avoid accidentally flipping labels during poisoning (clean-label assumption). 
{\em Secondly}, we run Algorithm~\ref{alg:alter_optim} to compute trigger pattern $\bM$ in the feature space. To increase realizability, we modify Algorithm~\ref{alg:alter_optim} to only consider features that have at least one mapped gadget for the trigger $\bM$. 
{\em Thirdly}, we run Algorithm~\ref{alg:final_trigger} to compute the trigger pattern in the problem space. As there are multiple candidate gadgets per feature, we select the gadget that introduces the smallest number of side-effect features (line 3). The final trigger  $\bM_p$ = $\bM$ + $\bmeta$ will include side-effect features after the set of gadgets $G$ is injected into the target apps.

\section{Evaluation: Conventional Backdoor Attack}
\label{sec:baseline}
As a baseline before evaluating our proposed attack, we first provide a quick evaluation of the stealthiness of the existing malware backdoor attack. We apply a recent defense method, MNTD, to the explanation-guided backdoor attack~\cite{explanation-backdoor}. 

\para{MNTD for Detecting Backdoors.}
MNTD assumes that backdoored models and clean models handle input queries differently, and the differences can be captured by a meta-classifier. Since the defender has no knowledge of the specific type of backdoor the attacker inserts, MNTD simply constructs a large number of ``shadow models'' where certain models are poisoned with {\em randomized} backdoors. Using these shadow models, MNTD trains a meta-classifier to detect whether a given model has been backdoored. The large number of randomly backdoored shadow models allows MNTD to generalize across different types of backdoor attacks (including those with previously unseen triggers), outperforming existing methods~\cite{mntd}. They also introduce a {\em query tuning} step, which co-optimizes the query inputs together with the meta-classifier to improve the detection performance.  

\para{Experiment Setup.}
We take the {\em stealthiest} version of the backdoor attack described by \citet{explanation-backdoor} (i.e., the ``greedy combined selection'' method). We set up a gradient-boosted decision tree (GBDT) classifier trained on the Ember PE malware dataset~\cite{ember} as the target model. 

For MNTD, we train a meta-classifier using 2,304 benign shadow models and 2,304 backdoored shadow models using jumbo learning on 2\% of the clean training set. 89\% of these shadow models are used for training and 11\% for validation. The backdoored shadow models are constructed using randomized triggers. Given the {\em realizability requirement}, we assume MNTD knows which features are modifiable.\footnote{
The set of modifiable features is common knowledge. In the Ember dataset, 2,316 out of the 2,351 features are created via feature hashing and thus are not directly modifiable. Among the 35 modifiable features, 17 are independently modifiable without affecting other features. 
}
For jumbo learning, we randomly pick features from 35 modifiable features to construct the trigger pattern and randomly set the feature values based on values observed in the 2\% training set. 
Other parameters of MNTD follow the default setting of MNTD.
After the MNTD meta-classifier is trained, we run it to classify 128 clean models and 128 backdoored models. The clean models are trained using a random sample of 50\% of the training set. The backdoored models are poisoned with the ``greedy combined selection'' method using 17 independently modifiable features (default setting), with a poison rate of 4\%.

\begin{table}[t]
\centering
\small
\begin{tabu}{l|c}
\tabucline[1.01pt]{-}
 \textbf{MNTD Configuration}    & \textbf{AUC (Avg $\pm$ Std)} \\  \hline
  MNTD w/o query tuning         &   0.800 $\pm$ 0.114    \\ %
  MNTD w/ query tuning          &   0.919 $\pm$ 0.052    \\ %
\tabucline[1.01pt]{-}
\end{tabu}
\vspace{-0.05in}
\caption{
 \textbf{MNTD Detection Result}---%
    \textmd{{\small 
    the detection AUC of MNTD against the explanation-guided backdoor attack.
    }}
}
\label{tab:mntd_on_exp_backdoor}
\vspace{-0.12in}
\end{table}

\para{Results.}
Table~\ref{tab:mntd_on_exp_backdoor} shows the results. We repeat the experiments 5 times, and report the average AUC (area under the ROC curve). AUC=1 indicates perfectly accurate detection while AUC=0.5 represents the results of random guessing. We observe that MNTD (with querying tuning) is highly effective in detecting backdoored models with an AUC of 0.919. 
The results suggest that, with a strong defense such as MNTD, the footprint of the realizable malware backdoor is conspicuous. 

\section{Evaluation: Jigsaw Puzzle Attack}
\label{sec:eval}

In this section, we evaluate our Jigsaw Puzzle (JP) attack. We start with a  ``feature-space'' attack to explore factors that affect the attack effectiveness and assess its detectability using recent defense methods. Later in \autoref{sec:eval-real}, we will move to the ``problem-space'' evaluation on realizable triggers. 

\subsection{Experiment Setup}
\label{sec:eval-setup}

\para{Dataset.}
We use an Android malware dataset sampled from AndroZoo~\cite{androzoo} between January 2015 and October 2016.\footnote{We focus on this time range because of the availability of malware family information. More recent malware lacks family information in AndroZoo~\cite{androzoo}.}
The apps are labeled following the same method used in prior works \cite{problem-space-sp20,tesseract}:  
an app is labeled ``benign'' if zero VirusTotal engines flagged it as malicious and is labeled ``malicious'' if at least four VirusTotal engines flagged it so. The rest is regarded as grayware (discarded). 
We sample proportionally to the total number of malware {\em each month} in AndroZoo with a sampling rate of 10\%. 
We use an adapted version of Drebin~\cite{drebin} to extract the feature vectors of these apps and train the binary malware classifier. We remove 396 (0.26\%) apps due to errors in feature extraction (e.g., invalid APK files). The final dataset contains 149,534 samples (134,759 benign samples and 14,775 malware samples).

To obtain the malware family information, we leverage Euphony~\cite{euphony} (developed by the AndroZoo team~\cite{androzoo}). In total, we have 400 malware families in the dataset. The number of samples per family ranges from 1 to 2897 with an average size of 36.94 and a standard deviation of 223.38. The top 13 families contribute to 80\% of the total malware samples (see Figure~\ref{fig:cdf} in the Appendix).

\begin{table*}[t]
\centering
\footnotesize
\begin{tabu}{r|c|c|c|c|c|c}
\tabucline[1.01pt]{-}
{\bf Target Set Family }  & {\bf \# of Samples}     & \textbf{Trigger Size} & $ASR(\bX_T^{\ast})$ & $ASR(\bX_R^{\ast})$ & $FPR(\bX_B^{\ast})$ & $F_1$(main) \\ 
\hline
Plankton            &  34    &  20          &  0.977              &  0.183                &   0.0005   &  0.927    \\
Mobisec             &  48    &  20          &  0.979              &  0.234                &   0.0002   &  0.927    \\
Adwo                &  60    &  34          &  0.810              &  0.282                &   0.0001   &  0.928    \\
Youmi               &  65    &  26          &  0.800              &  0.476                &   0.0000   &  0.928    \\
Cussul              &  117   &  23          &  0.916              &  0.663                &   0.0001   &  0.927    \\
Tencentprotect      &  142   &  23          &  0.954              &  0.500                &   0.0002   &  0.927    \\
Anydown             &  188   &  17          &  0.959              &  0.140                &   0.0004   &  0.924    \\
Leadbolt            &  210   &  18          &  0.927              &  0.087                &   0.0009   &  0.925    \\
Revmob              &  631   &  46          &  0.860              &  0.618                &   0.0000   &  0.925    \\
Airpush             &  1,021 &  47          &  0.742              &  0.123                &   0.0007   &  0.923    \\
\tabucline[1.01pt]{-}
\end{tabu}
\vspace{-0.05in}
\caption{
\textbf{Attack Results}---%
    \textmd{{\small 
     attack effectiveness in the feature space. The attacker aims for a high $ASR(\bX_T^{\ast})$, a low $ASR(\bX_R^{\ast})$, and a low $FPR(\bX_B^{\ast})$. The main task $F_1$ of the clean model is 0.926, which is comparable with the $F_1$(main) of the poisoned models in the table. 
    }}
}
\label{tab:attack_feature_space}
 \vspace{-0.12in}
\end{table*}

\para{Configurations.}
We randomly split the dataset for training (67\%) and testing (33\%). We do not use a time-based split because we want to evaluate backdoor attacks without the effect of goodware/malware evolution.
To improve training efficiency, we follow the suggestion from \citet{secsvm} to reduce the feature space. We use the LinearSVM $L_2$ regularizer to select the top 10,000 features---which maintains a similar accuracy as using the full feature set. Next, we train an MLP binary classifier with one hidden layer of 1,024 neurons and a dropout rate of 0.2. We use an MLP model because it has been successfully applied to malware classification in prior work~\cite{harang2020sorel20m,xu2018deeprefiner,explanation-backdoor,www21-mlp, chen2020training}. 

To run the JP attack, we first select a target family $T$ that the malware author aims to protect. Then we run Algorithm~\ref{alg:alter_optim} for at most 200 iterations to compute the trigger pattern $\bM$ and construct a poisoning set to train the target classifier. By default, we set a low {\em poisoning rate} of 0.1\% (i.e., the poisoning set is only 0.1\% of the original training set).  
As discussed before, we do not flip the labels of the poisoning samples (i.e., they keep their original ``benign'' label). We set batch size $M = 5$. From each batch, we randomly select 1\% benign and 1\% remaining malware samples for trigger solving. By default, we set $\lambda_1=5$ and set $\lambda_2=\lambda_3=v= 1$. We set $\lambda_1$ higher than the others in order to prioritize the protection of the target set malware (we can tolerate some accidental protection of the remaining malware). $\lambda_4$ is initialized as 0.001 to control the trigger size. To account for the randomness of training, we repeat the training process 5 times (with the same trigger $\bM$) and report the average results. 

\para{Evaluation Metrics.}
We evaluate our attack using different types of {\em test samples} on the poisoned model ($f^{\ast}$). We use $\bX$ to denote clean samples (without a trigger) and use $\bX^{\ast}$ to denote triggered samples. We consider four key metrics:

First, $ASR(\bX_T^{\ast})$ is the Attack Success Rate of the triggered target samples. It is the proportion of triggered malware samples in the target set $T$ that are classified as ``benign''. The attacker aims to obtain a high $ASR(\bX_T^{\ast})$ to evade detection. 

Second, $ASR(\bX_R^{\ast})$ is the Attack Success Rate of triggered remaining malware. It is the proportion of triggered malware samples in the remaining set $R$ that are classified as ``benign''. This metric measures how likely the trigger (accidentally) protects other malware families. The attacker aims to maintain a low $ASR(\bX_R^{\ast})$ to keep the attack stealthy. 

Third, $FPR(\bX_B^{\ast})$ is the False Positive Rate on triggered benign samples. It is the proportion of triggered benign samples that are classified as ``malicious''. The attacker aims to keep $FPR(\bX_B^{\ast})$ low (comparable to that of the clean model). 

Fourth, $F_1$(main) is the $F_1$ score on clean samples (which is usually referred as the ``main task'' performance~\cite{xie2020dba, bagdasaryan2020blind}). We use $F_1$ score instead of accuracy since our dataset is imbalanced. To avoid raising suspicion, the attacker aims for a high $F_1$(main) that is comparable to that of the clean model. 

Note that for the first three metrics, we only consider test samples that can be correctly classified by the {\em clean model}. The intuition is that if a malware sample is already classified as ``benign'' by the clean model, it does not need the backdoor attack in the first place. This allows us to explicitly measure the impact of the backdoor.  

\subsection{Attack Effectiveness}
\label{sec:eval-basic}
We start our evaluation in the feature space to understand important factors that affect the attack effectiveness. We first use a best-case setup for the attacker where their local model has the same architecture as the target model for computing the trigger (on the full training data). Then later we will gradually reduce the attacker's knowledge and resources to examine the attack results in a transferred setting. To show the attack is generally applicable to different malware families, we randomly select 10 families of different sizes as the target family $T$ to run the JP attack, as shown in Table~\ref{tab:attack_feature_space}. 

We have four important observations.
First, the attack is effective on different target families. For most families, the attack success rate on the target trigger samples ($ASR(\bX_T^{\ast})$) is above 0.9. A few families such as Mobisec and Plankton have an $ASR(\bX_T^{\ast})$ over 0.97. In each case, the attack has generally a much lower success rate on the ``remaining'' malware set ($ASR(\bX_R^{\ast})$), confirming the backdoor trigger is ``selective''.

Second, we show the trigger does not affect the benign samples, with an extremely low $FPR(\bX_B^{\ast})$. 
In the rest of the paper, we omit $FPR(\bX_B^{\ast})$ from the result tables for brevity, since it consistently stays at this low level. 
Third, the main task is not affected by the backdoor. The $F_1$ score of the main task (clean-sample classification) is always above 0.92, which is on par with the $F_1$ score of the clean model (0.926). 
Fourth, the trigger size is 10--50, which is within a reasonable range. The target classifier uses 10,000 features. On average a clean malware (benign) sample has 50.2 (49.5) features of non-zero values with a maximum of 211 (182) non-zero features. A trigger of this size should not raise anomalies.

In Table~\ref{tab:attack_feature_space}, we notice a few families do not perform as well as the others. For example, the large family Airpush (1,021 samples) has a slightly lower $ASR(\bX_T^{\ast})$ of 0.742. Cussul and Tencentprotect have a relatively higher $ASR(\bX_R^{\ast})$ (0.663 and 0.500). Later in \autoref{sec:eval-case}, we will use these as case studies and investigate ways to improve the performance. Note that $ASR(\bX_R^{\ast})$ of 0.5--0.6 does not mean the attack has failed. In our later evaluation (\autoref{sec:eval-defense1}), we find this $ASR(\bX_R^{\ast})$ is sufficient to remain stealthy against existing defenses (e.g., MNTD).

\subsection{Analyzing Impacting Factors}
\label{sec:factors}
Next, we restrict the attacker's knowledge and capability to explore key factors of the attacker's success. Due to the large number of experiments needed for this analysis, we select 3 families from Table~\ref{tab:attack_feature_space} for an in-depth analysis. Mobisec and Leadbolt represent two good-performing families with small (48) and large sizes (210); Tencentprotect represents an underperforming family with a slightly high $ASR(\bX_R^{\ast})$.

\begin{table}[t]
\centering
\resizebox{\columnwidth}{!}{
\begin{tabu}{r|r|c|c|c|c}
\tabucline[1.01pt]{-}
{\bf Rate ($r$)} & {\bf Target Set }  &  \textbf{Trg. Size}  & $ASR(\bX_T^{\ast})$ & $ASR(\bX_R^{\ast})$  & $F_1$(main) \\ 
\hline
        &   Mobisec                 &     14            &       0.950            &       0.194         &    0.927        \\ %
10\%    &   Leadbolt                &     6             &       0.750            &       0.019         &    0.927        \\
        &   Tencentprotect          &     40            &       0.494            &       0.215         &    0.928        \\ \hline
        &   Mobisec                 &     14            &       0.929            &       0.235         &    0.928        \\ %
20\%    &   Leadbolt                &     4             &       0.777            &       0.019         &    0.926        \\
        &   Tencentprotect          &     49            &       0.906            &       0.490         &    0.928        \\ 
\tabucline[1.01pt]{-}
\end{tabu}
}
\vspace{-0.05in}
\caption{
\textbf{Limited Data Access}---%
    \textmd{{\small 
the attacker only has access to $r$\% of the training set to solve the trigger. 
}}
}
\label{tab:limited_data}
\vspace{-0.1in}
\end{table}

\para{Limited Training Data.} 
We first restrict the attacker's access to the training data. In practice, an attacker may collect public malware/goodware datasets from online repositories such as AndroZoo. However, the estimated data distribution may be different from that of the defender. In this experiment, the attacker can only use 10\% and 20\% of the training set to compute the trigger. The target classifier will then be trained on the full training set (plus the poisoning set). As shown in Table~\ref{tab:limited_data}, the attack is still effective on Mobisec and Leadbolt. For Tencentprotect, while the 10\% setting starts to affect its performance, the attack is still effective under the 20\% access (comparable with Table~\ref{tab:attack_feature_space}, with 100\% access).

\para{Incorrect Model Architecture.}
The next experiment examines the impact of architecture differences between the target model and the attacker's local model. Recall that the target model uses MLP (10000-1024-1). Here, we let the attacker use a simpler local model (10000-32-1) to compute the trigger. As shown in Table~\ref{tab:transfer}, the attack is still effective. The mismatched model architecture causes small performance degradation on Mobisec and Leadbolt. Interestingly, for Tencentprotect, $ASR(\bX_T^{\ast})$ is reduced to 0.900 (from 0.954), but the $ASR(\bX_R^{\ast})$ is also reduced to 0.291 (from 0.500) for better stealth. We also test a local model with a more complex architecture (10000-2048-1). The transferred attack performance is still comparable to that using the same architecture. Overall, our backdoor attack is transferable in these settings.

\begin{table}[t]
\centering
\resizebox{\columnwidth}{!}{
\begin{tabu}{r|r|c|c|c|c}
\tabucline[1.01pt]{-}
{\bf Local Model}    &  {\bf Target Set}  &  \textbf{Trg. Size} & $ASR(\bX_T^{\ast})$ & $ASR(\bX_R^{\ast})$ & $F_1$(main) \\ 
\hline
        &   Mobisec                 &    21        &      0.950          &        0.387          &    0.928        \\ %
10000-32-1      &   Leadbolt                &    29        &      0.985          &        0.659          &    0.928     \\
        &   Tencentprotect          &    25        &      0.900          &        0.291          &    0.928     \\ \hline
        &   Mobisec                 &    22        &      0.992          &        0.246          &  0.928          \\ %
10000-2048-1    &   Leadbolt                &    24        &      0.947          &        0.206          &   0.927       \\
        &   Tencentprotect          &    24        &      0.968          &        0.494          &   0.927      \\
\tabucline[1.01pt]{-}
\end{tabu}
}
\vspace{-0.05in}
\caption{
\textbf{Transferred Attack}---%
    \textmd{{\small 
the attacker's local model has a different architecture from that of the target model (10000-1024-1). 
}}
}
\label{tab:transfer}
\vspace{-0.1in}
\end{table}

\para{Exposing Clean Target Set Samples to Defender.}
In practice, the defender may have previously collected clean samples from the target family (e.g., old variants). If the defender's training has included these clean samples (with correct malware label), it may counteract the influence of the poisoning. To evaluate this, we select \textasciitilde2/3 of the target set $T$ samples, and expose these {\em clean samples} (with ``malicious'' label) to the target model during training and poisoning. We report the results in Table~\ref{tab:allow-subset-in-train}. As expected, the $ASR(\bX_T^{\ast})$ is reduced due to exposure to the clean samples. However, the success rate is still higher than 0.88, indicating the attack can overcome the counter-effect of these clean samples.%

\begin{table}[t]
\centering
\resizebox{\columnwidth}{!}{
\begin{tabu}{r|c|c|c|c}
\tabucline[1.01pt]{-}
{\bf Target Set }  &  \textbf{Trigger Size} & $ASR(\bX_T^{\ast})$ & $ASR(\bX_R^{\ast})$ & $F_1$(main) \\ 
\hline
Mobisec                 &    21        &      0.996          &        0.238          &    0.927     \\ %
Leadbolt                &    25        &      0.881          &        0.343          &    0.928     \\
Tencentprotect          &    32        &      0.885          &        0.522          &    0.929     \\
\tabucline[1.01pt]{-}
\end{tabu}
}
\vspace{-0.05in}
\caption{
\textbf{Exposing Clean Target Samples to Defender}---%
    \textmd{{\small 
we expose 2/3 of the target set samples (clean samples with correct malware labels) to the target model. The attack is still effective.  
}}
}
\label{tab:allow-subset-in-train}
\vspace{-0.12in}
\end{table}

\subsection{Evaluating with Defense Methods}
\label{sec:eval-defense1}

To assess the attack's stealthiness, we run the attack against various defenses.

\para{Defense Methods.} For our evaluation, we select one input-level detection method: STRIP~\cite{strip}, one dataset-level defense: Activation Clustering (AC)~\cite{activationclustering}, and two model-level inspection methods: MNTD~\cite{mntd} and Neural Cleanse~\cite{neuralcleanse}. 

More specifically, STRIP~\cite{strip} detects triggered inputs by adding up (or \textit{mixing}) a given sample with many different clean samples. If this sample contains a trigger, then all of the mixed samples are likely to be classified to the same target label due to the trigger properties, leading to a low prediction entropy. AC~\cite{activationclustering} detects poisoned samples from the training dataset. The intuition is that clean samples and poisoned samples should show different neuron activation patterns, and AC looks for such differences in the last hidden layer. 
For model-level defenses, MNTD~\cite{mntd} is already introduced earlier in \autoref{sec:baseline}. Neural Cleanse~\cite{neuralcleanse} searches for possible trigger patterns that can cause the universal backdoor effect, i.e., classifying all triggered samples to the target label. 

Due to space limitations, our discussion below focuses on MNTD as we find it performs better than all other selected approaches (MNTD is also the most recent method). We also briefly discuss the results from STRIP, which shows some effectiveness on the baseline attack. Both AC and Neural Cleanse are ineffective against our selective backdoor attack (details are presented in Appendices \ref{sec:eval-ac} and \ref{sec:eval-nc}).

\para{Experiment Setting.} 
Considering most existing defenses are designed for image datasets and multi-class classifiers, we first check their {\em baseline performance} in our setting (sparse feature vectors for binary classification).
To do so, we run an experiment with a conventional ``universal'' backdoor attack. The trigger is non-selective, meaning {\em any} malware samples with the trigger will be classified as ``benign''. We implement this universal backdoor by selecting the top benign features as the trigger (features are ranked by the LinearSVM $L_2$ regularizer). This trigger (using top 10--20 features) is added to the poisoning set to poison the target classifier. We validate that this backdoor attack is effective with an ASR of 99.98\%.

After the baseline experiment, we then run our JP attack to examine the performance difference, which highlights the extra stealth introduced by our attack.

\begin{figure}[t]
\begin{subfigure}[t]{0.49\linewidth}
  \includegraphics[width=\linewidth]{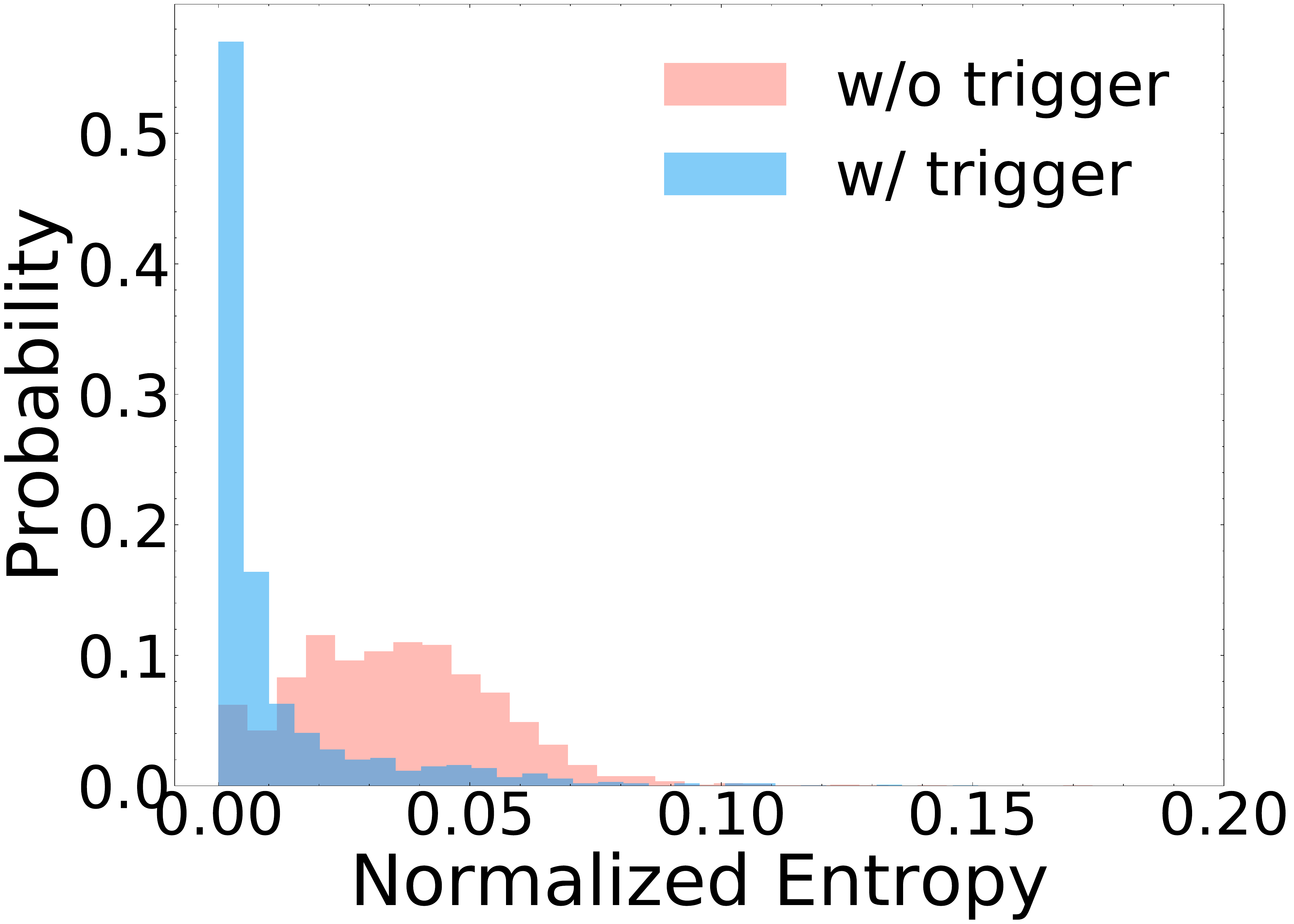}
  \vspace{-0.1in}
  \subcaption{Baseline}
  \label{fig:strip_baseline}
\end{subfigure}
\vspace{-0.05in}
\begin{subfigure}[t]{0.5\linewidth}
  \includegraphics[width=\linewidth]{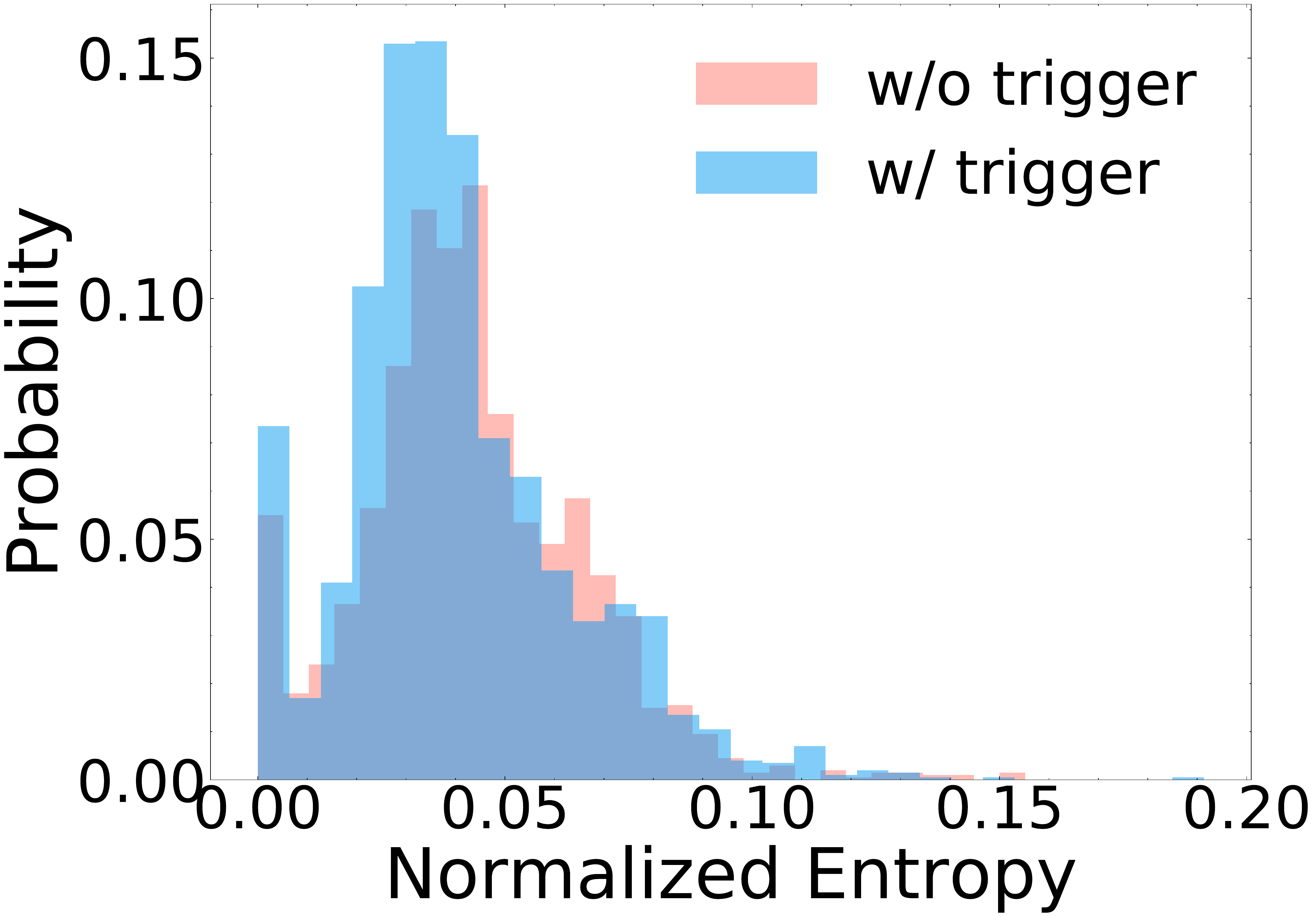}
  \vspace{-0.1in}
\subcaption{JP Attack (Mobisec)}
  \label{fig:strip_our}
\end{subfigure}
\vspace{-0.05in}
\caption{\textbf{STRIP: Entropy Histogram}---%
    \textmd{{\small 
   histogram of prediction entropy for 2000 clean and 2000 triggered samples. 
    }}
}
\label{fig:strip}
\end{figure}

\begin{table}[t]
\centering
\footnotesize
\begin{tabu}{@{}lccc@{}}
\tabucline[1.01pt]{-}

{\bf Attack Method}             & \textbf{False Reject.}      & \textbf{False Accept.}              & {\bf AUC}     \\ 
             & \textbf{Rate (FRR)}      & \textbf{Rate (FAR)}              &   ({\bf Avg $\pm$ Std}) \\ \hline
\multirow{2}{*}{Baseline Attack}       & 0.03              &  0.970 $\pm$ 0.021        & \multirow{2}{*}{0.801 $\pm$ 0.055}    \\
                                & 0.15              &  0.335 $\pm$ 0.141        &               \\ \hline \hline
\multirow{2}{*}{$T$=Mobisec}        & 0.03              &  0.970 $\pm$ 0.005        &  \multirow{2}{*}{0.486 $\pm$ 0.035}             \\
                                & 0.15              &  0.883 $\pm$ 0.032        &               \\ \hline
\multirow{2}{*}{$T$=Leadbolt}       & 0.03              &  0.972 $\pm$ 0.004        &  \multirow{2}{*}{0.396 $\pm$ 0.021}      \\
                                & 0.15              &  0.900 $\pm$ 0.012        &            \\ \hline
\multirow{2}{*}{$T$=Tencentprotect} & 0.03              &  0.980 $\pm$ 0.003        &  \multirow{2}{*}{0.472 $\pm$ 0.032}     \\
                                & 0.15              &  0.899 $\pm$ 0.017        &               \\
\tabucline[1.01pt]{-}
\end{tabu}
\vspace{-0.05in}
\caption{
\textbf{STRIP against Conventional and Selective Backdoor}---%
    \textmd{{\small 
 STRIP is moderately effective against the conventional baseline attack (AUC$=$0.801) but is ineffective against our selective backdoor attack (AUC$<$0.486). 
 }} 
}
\label{tab:strip_feature_space}
\vspace{-0.15in}
\end{table}

\subsubsection{STRIP Evaluation Results}
\label{sec:eval-strip}
For this evaluation, we follow the recommended setting of STRIP.\footnote{We obtained the code from \url{https://github.com/garrisongys/STRIP}; we validated the implementation with the CIFAR-10 dataset.} 
We randomly pick 2000 clean samples and 2000 triggered samples (containing both malware and benign examples). To classify whether a given sample is triggered, we mix this sample with one of the other 100 random clean samples to create 100 mixed vectors. Then we feed the vectors to the target classifier to calculate the prediction entropy. We repeat the experiments 5 times to report the average results. 

We find that STRIP shows some effectiveness on the baseline universal backdoor attack, but is ineffective against our attack. Figure~\ref{fig:strip} shows the histogram of prediction entropy obtained from clean and triggered inputs. 
Figure~\ref{fig:strip_baseline} shows the STRIP results on the baseline universal backdoor. We observe some separation between the clean and triggered inputs. 
As shown in Table~\ref{tab:strip_feature_space}, if we take a false rejection rate (FRR) of 15\% (classifying clean inputs as triggered), it produces a false acceptance rate (FAR) of 33.5\% (classifying triggered inputs as clean). The overall AUC is 0.801. This detection performance is slightly worse than that originally reported on image classifiers~\cite{strip}, possibly due to the binary-valued sparse feature vectors. When adding up two sparse vectors, it is easier to create out-of-distribution samples (which increases the prediction entropy even for triggered samples). 

In Figure~\ref{fig:strip_our}, we show the prediction entropy against our JP attack (selective backdoor). We observe the two entropy distributions overlap with each other and it is much more difficult to create separation. As shown in Table~\ref{tab:strip_feature_space}, when we set the FRR as 15\%, the FAR is 0.883 or higher (for all three target families). The overall AUC is below 0.486. The result confirms that STRIP is ineffective in detecting the JP attack.

\subsubsection{MNTD Evaluation Results}
\label{sec:eval-mntd}
For this evaluation, we use the original code of MNTD.\footnote{We obtained the code from \url{https://github.com/AI-secure/Meta-Nerual-Trojan-Detection}}
To effectively apply MNTD to our malware dataset, we have communicated with the authors of MNTD and configured MNTD based on their suggestions (see Appendix~\ref{sec:mntd-config} for details). 

To detect the {\em baseline attack} (i.e., universal backdoor), we train 2,304 clean shadow models and another 2,304 backdoored shadow models. We split these shadow models using 89\% for training and 11\% for validation. After training the MNTD meta-classifier, we use it to classify 256 clean models and 256 backdoored models. The 256 clean models are trained using a random sample of 50\% of the training set. The 256 backdoored models are poisoned by a universal backdoor that aims to misclassify {\em any} triggered malware samples as ``benign''. We assume MNTD has some knowledge about the top features (but does not know the exact features of the trigger). As such, the defenders randomly pick the top $n$ benign features (5$\leq n<$100) to create random triggers (jumbo learning on 2\% of the training set) to train MNTD. 

As shown in Table~\ref{tab:mntd_feature_space}, MNTD is highly effective against the universal backdoor (baseline) with an AUC of 0.960 when query tuning is enabled. This confirms that MNTD is at least applicable to our dataset and the binary classification setting. 

Next, we further evaluate MNTD against our JP attack. The configuration is mostly consistent with the above. However, since the JP attack is not restricted to using the top benign features, MNTD trained with top benign features does not perform well. As such, we allow MNTD to select any features (random $n$ features, 5$\leq n<$100) for the jumbo learning. Other parameters follow MNTD's recommended settings. We construct a testing set of 256 clean models and 256 backdoored models (with a poisoning rate between 0.1\% and 0.2\%).

As shown in Table~\ref{tab:mntd_feature_space}, our attack can evade the detection of MNTD. 
The detection AUCs are below 0.557 (barely better than random guessing) for all three target families. Importantly, we confirm that the selective backdoor attack on Tencentprotect can evade MNTD. Recall that Tencentprotect is considered an underperforming family because its $ASR(\bX_R^{\ast})$ (attack success rate on the remaining malware) is moderately high (0.500). As a sanity check, we also run the MNTD experiment for another high-$ASR(\bX_R^{\ast})$ family called ``Cussul'' with $ASR(\bX_R^{\ast})$=0.663. We confirm that the selective backdoor of Cussul can also evade MNTD. The results suggest that an $ASR(\bX_R^{\ast})$ around 0.5 to 0.6 can already provide sufficient stealth against existing detectors.

\begin{table}[t]
\centering
\footnotesize
\begin{tabu}{l|l|c}
\tabucline[1.01pt]{-}
 \textbf{MNTD Configuration}                     & \textbf{Attack Method}     & \textbf{AUC (Avg $\pm$ Std)} \\ 
\hline
                            &   Baseline attack &   0.836 $\pm$ 0.090     \\\cline{2-3}
                            &   $T$=Mobisec         &   0.544 $\pm$ 0.062    \\
  MNTD w/o query tuning     &   $T$=Leadbolt        &   0.557 $\pm$ 0.033    \\ %
                            &   $T$=Tencentprotect  &   0.508 $\pm$ 0.025    \\ \hline
                            \hline
                            &   Baseline attack         &   0.960 $\pm$ 0.077    \\\cline{2-3}
                            &   $T$=Mobisec         &   0.518 $\pm$ 0.027    \\
  MNTD w/ query tuning      &   $T$=Leadbolt        &   0.545 $\pm$ 0.035    \\ %
                            &   $T$=Tencentprotect  &   0.533 $\pm$ 0.032    \\ 
\tabucline[1.01pt]{-}
\end{tabu}
\vspace{-0.04in}
\caption{%
\textbf{MNTD against Conventional and Selective Backdoor}---%
    \textmd{{\small 
 MNTD (w/ query tuning) is highly effective against the conventional baseline attack (AUC$=$0.960), but is ineffective against our selective backdoor attack (AUC$<$0.557). 
 }}
}
\label{tab:mntd_feature_space}
\vspace{-0.1in}
\end{table}

\subsection{Case Study on Underperforming Families}
\label{sec:eval-case}

\begin{table}[t]
\centering
\small
\begin{tabu}{r|cc|c}
\tabucline[1.01pt]{-}
{\bf Target Set }  &   $ASR(\bX_T^{\ast})$ & $ASR(\bX_R^{\ast})$ & \textbf{Regres. Error} \\ 
\hline
Cussul                & 0.916 & 0.663  &   0.0313        \\ %
Tencentprotect        & 0.954 & 0.500  &   0.0088        \\
\hline
Mobisec               & 0.979 & 0.234  &  0.0006        \\
Leadbolt              & 0.927 & 0.087  &  0.0010        \\
\tabucline[1.01pt]{-}
\end{tabu}
\caption{
\textbf{Regression Analysis}---%
    \textmd{{\small 
    we run a regression to separate the target family ($T$) and the remaining set ($R$), and report the regression error. A larger error indicates $T$ and $R$ are harder to separate.
}}
}
\label{tab:logistic}
\vspace{-0.17in}
\end{table}

To understand the reasons behind the underperforming families in the main experiment, we perform several case studies. 

\para{Cussul \& Tencentprotect.} As shown in Table~\ref{tab:attack_feature_space}, Cussul and Tencentprotect have a high success rate on the target set but their $ASR(\bX_R^{\ast})$ is higher (0.500--0.663) than other families. 
Although our evaluation has shown that their $ASR(\bX_R^{\ast})$ is sufficient to evade existing detectors  (\autoref{sec:eval-defense1}), we still would like to understand the reason behind their high $ASR(\bX_R^{\ast})$. 

After analyzing the feature distributions of these two families, we observe that the common features of Cussul and Tencentprotect are also common in the remaining malware. In other words, we suspect that the target malware samples  ($T$) in Cussul and Tencentprotect are too similar to the remaining malware ($R$), making it difficult to find a trigger that selectively protects $T$ while ignoring $R$. 

To validate this hypothesis, we run a simple logistic regression analysis, attempting to separate the target set $T$ and $R$. As shown in Table~\ref{tab:logistic}, for Cussul, the analysis returns a relatively large regression error (0.0313) which is similarly high for Tencentprotect. This confirms that their common characteristics with other malware make it hard to separate them from the remaining set. Note that this high similarity could be caused by the feature engineering---a different feature engineering method might mitigate this issue. For comparison, we run the regression analysis for two well-performing families, Mobisec and Leadbolt. Both return much lower regression errors (0.0006 and 0.0010), meaning they can be more easily separated from the remaining families, so that it is easier to create a selective backdoor for them.

\para{Airpush.} Airpush is a large family with 1,021 samples. As shown in Table~\ref{tab:attack_feature_space}, Airpush's $ASR(\bX_R^{\ast})$ is reasonably low (0.123). However, its success rate on the target set $T$ is among the lowest ($ASR(\bX_R^{\ast})$=0.742). We analyze the failed Airpush samples and find that they usually carry a large number of malicious features. It is possible that the small trigger is insufficient to overturn the ``malicious'' label. To further improve its $ASR(\bX_R^{\ast})$, we slightly tune the corresponding hyperparameters in the loss function that control $ASR(\bX_R^{\ast})$. 
For instance, by increasing $\lambda_1$ to 10 (from 5) and $\lambda_2$ to 2 (from 1) while keeping $\lambda_3=1$, we can get an $ASR(\bX_T^{\ast})$ of 0.908 and an $ASR(\bX_R^{\ast})$ of 0.423, which is on par with other families.

\begin{table}[t]
\centering
\resizebox{\columnwidth}{!}{
\begin{tabu}{r|c|c|c|c}
\tabucline[1.01pt]{-}
{\bf Target Set }  &  \textbf{Trg. Size ($\bM_{p}$)} & $ASR(\bX_T^{\ast})$ & $ASR(\bX_R^{\ast})$   & $F_1$(main) \\ 
\hline
Mobisec                 &    31        &      0.925          &        0.133          &    0.926     \\ %
Leadbolt                &    6         &      0.791          &        0.041          &    0.926     \\
Tencentprotect          &    53        &      0.920          &        0.418          &    0.926     \\
\tabucline[1.01pt]{-}
\end{tabu}
}
\vspace{-0.05in}
\caption{
\textbf{Attack Results (Problem Space)}---%
    \textmd{{\small 
Selective backdoor attack in the problem space. Results are comparable with the feature-space attack (Table~\ref{tab:attack_feature_space}). 
}}
}
\label{tab:attack_problem_space}
\vspace{-0.13in}
\end{table}

\section{Problem-Space Attack and Defense}
\label{sec:eval-real}

In this section, we extend our evaluation to the problem-space by realizing the triggers in the malware/benign software code. Following the methodology described in \autoref{sec:real}, we first extract the mapping between features and benign gadgets. Out of the 10,000 features, we are able to extract gadgets for 2,171 features using the enhanced harvesting tool. For certain features, we cannot extract the corresponding gadgets due to implementation limitations of FlowDroid~\cite{arzt2014flowdroid} that serves as the core instrumentation library for the harvesting tool. While the feature coverage can be further improved (with additional engineering efforts), we believe this mapping is sufficient for proof-of-concept. Based on the mapping, we run the problem-space attack by considering the side-effect features. 

The additional computational overhead introduced by the problem-space attack is acceptable. While the gadget harvesting process can be time-consuming (144 hours, using a commodity server), we argue it is a {\em one-time effort}. Once the gadget-feature mapping is extracted, it can be reused to run {\em any} future JP attacks. With a database of gadgets, it only takes several minutes to compute the final trigger with the feature-space trigger. Further details are presented in Appendix~\ref{sec:overhead}.

\para{Attack Effectiveness.} 
Table~\ref{tab:attack_problem_space} shows the attack results in the problem space. We find that the attack is still effective using realizable triggers. For Mobisec, the $ASR(\bX_T^{\ast})$ is still high (0.925) with an $ASR(\bX_R^{\ast})$ of 0.133. The attack becomes slightly weaker on Leadbolt with an $ASR(\bX_T^{\ast})$ around 0.8 but still a low $ASR(\bX_R^{\ast})$ of 0.041. The weakened attack is likely due to the much smaller trigger size (6 features). The final trigger is small because not all candidate features in the original trigger have a mapping gadget. 
We also observe that Tencentprotect has a comparable result with slightly increased trigger size (from 23 to 53) due to the side-effect features.    

\para{Evaluation against MNTD.} To assess the stealthiness of the realizable triggers, we again use MNTD following the same setting of \autoref{sec:eval-defense1}. Since other defenses such as STRIP, AC, and Neural Cleanse are easier to evade (as shown in \autoref{sec:eval-defense1}), we only present the strongest defense (MNTD) here for brevity. 
The detection results are presented in Table~\ref{tab:mntd_problem_space_jumbo_10000}. We find that the selective backdoor attack still successfully evades MNTD in the problem space---regardless of whether query tuning is enabled or not, the detection AUC is barely above 0.5. 

Next, we aim to further help MNTD by giving away the exact list of 2,171 features for which the attacker can harvest gadgets. Note that the list is highly dependent on the attacker's gadget harvesting strategies and the benign applications used for the harvesting. While it is unrealistic that the defender knows the exact list, we want to see if such information can help MNTD.  Table~\ref{tab:mntd_problem_space_jumbo_2171} shows the evaluation results which demonstrate that the selective backdoor attack can still evade the detection of MNTD, even if we assume the defender knows the exact list of realizable features.

\begin{table}[t]
\centering
\small
\begin{tabu}{l|c|c}
\tabucline[1.01pt]{-}
 \textbf{MNTD Configuration}                     & \textbf{Target Set}     & \textbf{AUC (Avg $\pm$ Std)} \\ 
\hline
                            &   Mobisec         &   0.524 $\pm$ 0.039    \\
  MNTD w/o query tuning     &   Leadbolt        &   0.533 $\pm$ 0.032    \\ %
                            &   Tencentprotect  &   0.566 $\pm$ 0.088    \\ \hline
                            &   Mobisec         &   0.524 $\pm$ 0.019    \\
  MNTD w/ query tuning      &   Leadbolt        &   0.514 $\pm$ 0.017    \\ %
                            &   Tencentprotect  &   0.521 $\pm$ 0.037    \\ 
\tabucline[1.01pt]{-}
\end{tabu}
\caption{
\textbf{MNTD Detection (Problem Space)}---%
    \textmd{{\small 
MNTD against selective backdoor in the problem space; MNTD jumbo learning constructs randomized triggers with all features. 
 }}
}
\label{tab:mntd_problem_space_jumbo_10000}
\end{table}

\begin{table}[t]
\centering
\small
\begin{tabu}{l|c|c}
\tabucline[1.01pt]{-}
 \textbf{MNTD Configuration}                     & \textbf{Target Set}     & \textbf{AUC (Avg $\pm$ Std)} \\ 
\hline
                            &   Mobisec         &   0.529 $\pm$ 0.033    \\
  MNTD w/o query tuning     &   Leadbolt        &   0.532 $\pm$ 0.026    \\ %
                            &   Tencentprotect  &   0.556 $\pm$ 0.080    \\ \hline
                            &   Mobisec         &   0.515 $\pm$ 0.009    \\
  MNTD w/ query tuning      &   Leadbolt        &   0.476 $\pm$ 0.021    \\ %
                            &   Tencentprotect  &   0.490 $\pm$ 0.021    \\ 
\tabucline[1.01pt]{-}
\end{tabu}
\caption{
\textbf{MNTD Detection (Problem Space)}---%
    \textmd{{\small 
 MNTD against selective backdoor in the problem space; MNTD constructs randomized triggers with 2,171 realizable features only.
 }}
}
\label{tab:mntd_problem_space_jumbo_2171}
\vspace{-0.14in}
\end{table}

\section{Discussion}
\label{sec:discuss}

\vspace{-0.05in}

\para{Lessons learned.}
There are multiple explanations behind the stealthiness of our JP attack against existing defenses. First and foremost, the selective backdoor design has reduced the footprint of the backdoor even within the same class. This breaks existing defenses that have assumed any triggered samples (within a class) will be misclassified to the target label. We show that MNTD, which works well on conventional backdoors in malware classifiers, can be evaded by the selective backdoor.   
Second, some defenses (e.g., STRIP) might be suitable for image data but are not optimized for malware samples (sparse feature vectors). Defense techniques that are designed for multi-classification models (e.g., Neural Cleanse, Appendix~\ref{sec:eval-nc}) also suffer when used on binary classifiers. Finally, our problem-space trigger is not limited to independently modifiable features, which also helps improve stealthiness by increasing the search space for defenders.

\para{Ideas for Countermeasures.}
While designing a new adaptive defense is out of scope of this paper, we here discuss potential directions. To defend against a selective backdoor, existing defenses need to revisit their assumptions as the attackers may target only a subset of a class. An adaptive defense must make a guess on which subset is the target. A na\"ive defense may take one malware family at a time and exhaustively scan for a selective backdoor in each family. However, the attacker can evade this defense by dividing their malware family into sub-families and designing a different selective backdoor for each. Additionally, the malware author can  disregard the old samples that are already detected/fingerprinted by AV engines and focus on protecting new variants to be disseminated in the future. By increasing the difference between new and old variants, the selective backdoor for the new variants will be more difficult to detect.

Another defense idea is inspired by the observations from our case studies in~\autoref{sec:eval-case}. We have shown that if a malware family is too ``generic'' (with a high similarity to the remaining malware families), it is more difficult to create a selective backdoor. Therefore, defenders might improve the feature engineering process to increase the data homogeneity within the ``malware'' class. This can be done by further removing some family-specific features (to reduce selective backdoor risk) while preserving key malware features (to maintain the performance of the malware detection task).

\begin{table*}[t]
\centering
\scriptsize
\begin{tabu}{rccccl}
\toprule 
{\bf Method}       & {\bf Control}         & {\bf Clean} & {\bf Application}    & {\bf Problem} & {\bf Attack Effect}\\
                & {\bf Train. Process}  & {\bf Label} & {\bf Domain} & {\bf Space}   &  $S$: sample; $T$: trigger; $L$: Label; $C$: Class \\
\midrule 
BadNets\cite{gu2017badnets}  & \CIRCLE & \Circle & IMG & NA & Any $S$ + uniform $T$ $\rightarrow$ target $L$ \\
Dynamic backdoor \cite{salem2020dynamic}  & \CIRCLE & \Circle & IMG & NA & Any $S$ + dynamic $T$ $\rightarrow$ one/more target $L$ \\
Invisible backdoor \cite{li2021invisible}  & \CIRCLE & \Circle & IMG & NA & Any $S$ + sample-specific $T$  $\rightarrow$ target $L$ \\
DFST \cite{cheng2020deep}  & \CIRCLE & \Circle & IMG & NA & Any $S$ + style-transfer $T$  $\rightarrow$ target $L$ \\
Composite attack \cite{lin2020composite}  & \CIRCLE & \Circle & IMG/Text & NA & Mixing $S \in C_A$ and $S \in C_B$ $\rightarrow$  target $L$\\  
Latent backdoor \cite{yao2019latent}  & \CIRCLE & \Circle & IMG & NA &  Any $S$ + uniform $T$ $\rightarrow$ target $L$ (student model only) \\

\midrule 

TaCT \cite{tang2021demon}  & \Circle & \Circle & IMG & NA &  $S \in C_A$ + uniform $T$ $\rightarrow$ target $L$ \\
WaNet \cite{nguyen2021wanet}  & \Circle & \Circle & IMG & NA & Any $S$ + image warping $T$ $\rightarrow$ target $L$ \\
Subpopulation attack \cite{jagielski2021subpopulation} & \LEFTCIRCLE  & \Circle & IMG/Text/Tabular & NA & Subpopulation sample $\rightarrow$ non-origin $L$ \\

\midrule 

Clean-label backdoor \cite{turner2018clean}  & \Circle & \CIRCLE & IMG & NA & Any $S$ + uniform $T$ $\rightarrow$ target $L$ \\
Reflection backdoor \cite{liu2020reflection}  & \Circle & \CIRCLE & IMG & NA & Any $S$ + image reflection $T$ $\rightarrow$ target $L$ \\
Poison frogs \cite{shafahi2018poison}  & \LEFTCIRCLE  & \CIRCLE & IMG & NA & Target $S$ $\rightarrow$  target $L$ \\

\midrule 

Exp-guided backdoor \cite{explanation-backdoor}  & \Circle & \CIRCLE & {\bf Malware} & \CIRCLE & $S \in C_A$ + uniform $T$ $\rightarrow$ target $L$ \\
Selective backdoor (Ours) & \Circle & \CIRCLE & {\bf Malware} & \CIRCLE & $S \in$ subset of $C_A$ + subset-specific $T$ $\rightarrow$ target $L$ \\

\bottomrule 
\end{tabu}
\vspace{-0.05in}
\caption{
\textbf{Related Backdoor Attacks Focusing on Stealth}---%
    \textmd{{\small 
    ``Control Training Process'' means the attacker trains the backdoored model with full knowledge and control. (\LEFTCIRCLE) denotes methods that require knowledge of target models' loss function or model architecture. Compared with the existing attack \cite{explanation-backdoor}, our attack further improves the stealthiness to evade detectors such as MNTD.
 }}
}
\label{tab:related}
\vspace{-0.14in}
\end{table*}

\para{Limitations.}
Our study has a few limitations. First, our evaluation is mainly based on an Android malware dataset. This is because it would require extensive engineering efforts to develop a new gadget harvesting tool for other binary types (e.g., PE files). Since gadget extraction (binary analysis) is not the main focus of the paper, we use the Android malware as a proof-of-concept for our idea. Second, our main experiment simply uses one set of hyperparameters for all malware families. It is possible that further tuning the hyperparameters for each family may produce better results (future work).   
Finally, there is still room to make the attack even stealthier (as discussed above). We leave further experiments on adaptive attacks against new countermeasure ideas to future work.

\section{Related Work}
\label{sec:related}

We discuss existing works that aim to make backdoor attacks {\em stealthier}, and categorize them under different threat models.  Table~\ref{tab:related} highlights a subset of representative works.    

\para{Attacker Controlled Training.}
In the canonical supply chain backdoor attack, the adversary is assumed to control the training process to insert a backdoor (\eg BadNets~\cite{gu2017badnets}) and can arbitrarily label training examples (in contrast to clean-label attacks). Under this threat model, researchers have explored how to improve stealthiness by using dynamic triggers (\eg without fixing the trigger size or location)~\cite{salem2020dynamic}, creating sample-specific triggers using jointly trained encoders~\cite{li2021invisible, nguyen2020input}, or using ``image styles'' as triggers~\cite{cheng2020deep}.
Recently, the composite attack has been proposed~\cite{lin2020composite}, which mixes two examples from different classes to produce a trigger. The latent backdoor attack focuses on transfer learning~\cite{yao2019latent}---the attacker trains a teacher model containing an incomplete trigger; after the victim trains a student model using this teacher model, the trigger is completed and the backdoor effect activates. Other attacks under this threat model are those that manipulate image encoders of self-supervised learning models~\cite{jia2022badencoder}, insert backdoors into the latent space~\cite{doan2021backdoor}, exploit transformation functions~\cite{doan2021lira}, and directly edit the weights of vulnerable neurons~\cite{rakin2020tbt, chen2021proflip, liu2018trojaning}. 
In contrast to our attack, these methods give the attacker privileged control over the training process, and many do not generalize beyond the image domain (\eg style transfer).

\para{Attacker Controlled Data and Labeling.} An alternative threat model does not allow the attacker to control the training process itself, but only to provide poisoned data and labels. To increase stealthiness, some techniques exploit properties of image classification: TaCT~\cite{tang2021demon} uses triggers that only work for a given class and  WaNet~\cite{nguyen2021wanet} uses image warping as a trigger such that the trigger is imperceptible to humans. A recent subpopulation attack~\cite{jagielski2021subpopulation} does not use triggers, but instead supplies poisoned data targeting a specific ``subpopulation'' within the dataset. After training, the poisoned classifier will exclusively misclassify the target subpopulation. However, all these attacks still require that the attacker controls the labeling process to provide incorrect labels for the poisoned data. 

\para{Clean-Label Attacks.} 
Clean-label attacks do not require the attacker to control the labeling process~\cite{turner2018clean}. As such, the supplied poisoned data will have their original labels---this is the assumption in our work. An image-specific example is the reflection attack~\cite{liu2020reflection}, which creates natural-looking triggers by applying the reflection effect from glasses and windows to everyday objects. Poison frog attacks~\cite{shafahi2018poison} aim to misclassify {\em one specific example}. Rather than triggers they use specifically crafted, clean-labeled data to poison the model, however, the attacker must know the target model's loss function to compute the special poisoning data. Batch-Order Backdoor (BOB) attacks~\cite{shumailov2021manipulating} create a backdoor by changing the order of training examples that are fed into the model. 
Our proposed backdoor attack is also a clean-label attack, however, with a specific focus on stealthier backdoors for malware classifiers.

\para{Backdooring Malware Classifiers.} 
Most existing backdoor attacks cannot be applied to malware classifiers because (1) the techniques are specifically designed for images (\eg style transfer, reflection effect) and/or (2) the trigger computation cannot be easily realized in the problem space. Existing works targeting malware classifiers~\cite{explanation-backdoor, li2021backdoor} focus on conventional backdoors that aim to misclassify arbitrary malware samples. In contrast, we have shown that a \textit{selective backdoor} improves stealthiness, following the intuition that a malware author would prioritize protecting their own malware family instead of all malware in general. Furthermore, \citet{li2021backdoor} still require the attacker to flip the label.

\section{Conclusion}
\label{sec:conl}

In this paper, we empirically evaluate the stealthiness of existing backdoor attacks in malware classifiers and show their detectability. To improve stealth, we propose Jigsaw Puzzle (JP), a selective backdoor attack that aims to exclusively protect a malware author's samples while ignoring other malware. We validate this idea in both the feature space and the problem space, against a series of defense methods such as MNTD, STRIP, AC, and NC. Our future work will look into effective defense methods against selective backdoor attacks.

\newpage

\begin{small}
\bibliographystyle{abbrvnat}
\bibliography{main}
\end{small}

\appendix

\section{Evaluation with Activation Clustering}
\label{sec:eval-ac}

\begin{table}[ht]
\centering
\resizebox{\columnwidth}{!}{
\footnotesize
\begin{tabu}{@{}lc|cc|cc@{}}
\tabucline[1.01pt]{-}
\multirow{2}{*}{\bf Target Set}         &  {\bf Model}                     &    \multicolumn{2}{c|}{\bf Benign}                                        &     \multicolumn{2}{c}{\bf Malware}                                       \\ 
                                        & \textbf{Type}   & \textbf{Size}     &   {\bf Silhouette}  &     \textbf{Size}    & {\bf Silhouette}     \\ \hline
\multirow{2}{*}{Mobisec}                & Clean           &  (0.43, 0.57)     &     0.11            &     (0.04, 0.96)     &   0.36               \\
                                        & Poison &  (0.30, 0.70)     &     0.13            &     (0.04, 0.96)     &   0.34               \\ \hline
\multirow{2}{*}{Leadbolt}               & Clean           &  (0.31, 0.69)     &     0.13            &     (0.04, 0.96)     &   0.34               \\
                                        & Poison &  (0.26, 0.74)     &     0.11            &     (0.04, 0.96)     &   0.34               \\ \hline
\multirow{2}{*}{Tencentprotect}         & Clean           &  (0.32, 0.68)     &     0.12            &     (0.04, 0.96)     &   0.35               \\
                                        & Poison &  (0.44, 0.56)     &     0.11            &     (0.04, 0.96)     &   0.34                \\ \hline
\tabucline[1.01pt]{-}
\end{tabu}
}
\vspace{-0.05in}
\caption{
\textbf{Activation Clustering against Selective Backdoor}---%
    \textmd{{\small 
    AC's implementation assumes that any cluster size smaller than 0.35 can be deemed as poisoned. A high silhouette score (above 0.10--0.15) also indicates the class is poisoned. 
    In our attack, only 0.1\% of benign samples are poisoned. We do not poison any malware.
 }} 
}
\label{tab:ac_feature_space}
\end{table}

Activation Clustering~\cite{activationclustering} aims to detect poisoning samples in the training set. The intuition is that clean and poisoning samples will have different activation patterns in the last hidden layer of the deep neural network. More specifically, the activations of clean samples will capture features related to their original class. However, the activations of poisoning samples will capture features related to its source class and also the trigger. As a result, if a given class contains poisoning samples, these samples' activation patterns can be clustered into two distinct groups (one represents poisoning samples, and the other represents clean samples). Since we assume poisoning samples would only take a small portion of the training set, the two clusters would have uneven sizes.   

We use the latest code of Activation Clustering (AC) provided by the authors\footnote{\url{https://github.com/Trusted-AI/adversarial-robustness-toolbox}} to evaluate our selective backdoor attack. For each class in the training set, the algorithm first obtains the activation values of the last hidden layer (1024 neurons) for all samples in the class. Then it reduces the vector dimensions from 1024 to 10 using Independent Component Analysis. Finally, it runs a K-means algorithm (K=2) on these vectors to separate them into two clusters for further analysis. 

AC determines the existence of backdoor by analyzing the cluster sizes and their silhouette score. First, AC flags a class as poisoned if it produces two highly uneven sized clusters, i.e., if the {\em relative size} of either cluster is smaller than $t$, the class is poisoned. In the AC's implementation, $t$ is set to $0.35$. Second, it looks into the tightness of the two clusters. If the clusters are tight (i.e., silhouette score of 0.10--0.15 or above), then it means the two clusters contain highly distinct patterns (i.e., poisoned). Otherwise, it means the two clusters are hard to separate (i.e, not poisoned). Note that we do not use their exclusionary reclassification analysis because it is designed for label-flipping attacks (our attack is clean-label). 

We run our selective backdoor attack against AC (using the same configuration as in \autoref{sec:eval-defense1}). We poison 0.1\% of the benign set, and we do not poison any malware samples. We run the experiments with three target families; for each target family, we train a clean model and a poisoned model. 

As shown in Table~\ref{tab:ac_feature_space}, AC does not work well on our selective backdoor attack. More specifically, there is not enough separation between ``clean'' and ``poisoned'' activation vectors. For cluster sizes, if we use AC's threshold $t=0.35$, then the entire malware class would be determined as poisoned (although we in fact do not poison any malware samples). At the same time, for the benign class, some of the clean models (Mobisec and Leadbolt) will be incorrectly determined as poisoned. If we further examine the silhouette score, we find that the scores are very close to the threshold values (0.10--0.15) regardless of whether the model is poisoned. Also, poisoned models do not necessarily have a higher silhouette score. 

Overall, the results suggest the selective backdoor is stealthy against AC. We suspect three possible reasons. First, AC assumes the label of the poisoned data has been manipulated/flipped to the target label. In our case, we keep the original label (``benign'') for the poisoning samples. Second, the selective backdoor may have reduced the differences in the activation patterns between clean and poisoning samples. Third, the dataset contains highly diverse samples even within the same class (for both goodware and malware). It breaks AC's assumption that clean samples within the same class are hard to separate.

\section{Evaluation with Neural Cleanse}
\label{sec:eval-nc}

\begin{table}[h]
\centering
\small
\begin{tabu}{@{}l|cc|cc@{}}
\tabucline[1.01pt]{-}
\multirow{2}{*}{\bf Target Set}  & \multicolumn{2}{c|}{\bf Benign}  & \multicolumn{2}{c}{\bf Malware} \\ 
                       & Clean  &    Poisoned                            &     Clean  &    Poisoned    \\ \hline
Mobisec                &  21    &     28  &                              6            &    6      \\
Leadbolt               &  21    &     20  &                              6            &    7      \\
Tencentprotect         &  21    &     22  &                              7            &    8      \\
\tabucline[1.01pt]{-}
\end{tabu}
\vspace{-0.05in}
\caption{
\textbf{Neural Cleanse against Selective Backdoor}---%
    \textmd{{\small 
    we present the trigger size inferred by NC from clean and poisoned models. There is no clear difference in the trigger sizes between clean and poisoned models, i.e., our attack is stealthy against NC. 
 }} 
}
\label{tab:nc_feature_space}
\end{table}

Neural Cleanse (NC)~\cite{neuralcleanse} is designed to search for a small perturbation (i.e., the trigger pattern) that allows any samples from all classes to be unanimously classified to the target label. NC is originally designed for multi-class classification models. It tries to infer a trigger for each of the classes---any class that has an anomalously small trigger is likely to be poisoned. The anomalously small trigger is determined by an outlier detection algorithm~\cite{neuralcleanse}. For this reason, NC is more suitable for a multi-class classification setting to run the outlier detection. If there are only two classes (i.e., binary classifier), it is more difficult to determine the outlier. We have attempted to adapt NC for binary classifiers after communicating with the authors of NC. 

More specifically, we start with the original code of NC\footnote{\url{https://github.com/bolunwang/backdoor}}, and modify the trigger injection method. The original injection method is designed for images: $A(\bx, \bM, \bm{\triangle}) = (1 - \bM_{i, j}) \cdot \bX_{i, j, c} + \bM_{i, j} \cdot \bm{\triangle}_{i, j, c}$ where $\bX$ is the original clean image. $\bm{\triangle}$ is the trigger pattern and $\bM$ is a 2D matrix deciding how much the trigger can overwrite the original image.
We change the trigger injection to $A(\bx, \bM) = (1 - \bM) \cdot x + \bM$ where $\bM$ is the reversed trigger. We convert $\bM$ to binary values with a value larger than 0.5 as 1 otherwise 0. When $\bM_i = 1$, the final feature value would be 1 regardless of the original feature value. While we keep the original feature value if $\bM_i = 0$. With this adapted generic form, we only allow adding a feature to the vector without any feature removal (to mimic our attack algorithm). We also change the learning rate from 0.1 to 0.001 and initialize the cost of the regularization term as 0.001 instead of 0. 
Other parameters follow the same setting as NC.

We run our selective backdoor attack (similar to \autoref{sec:eval-defense1}). We run the experiments with three different target families, and apply NC to infer triggers for both clean and poisoned models. Since we cannot run outlier detection on two classes (as described above), we simply report the inferred trigger size as NC takes the ``benign'' and ``malware'' as the target class, respectively. We want to see if there is a clear difference between the trigger size distribution inferred from the clean model and the poisoned model. The results are reported in Table~\ref{tab:nc_feature_space} with all trigger success rates above 0.99. 

From Table~\ref{tab:nc_feature_space}, we observe that there is no clear difference in the trigger size distribution between the {\em clean model} and the {\em poisoned model}. This means NC cannot effectively determine whether a model is poisoned based on the trigger size information. We suspect that the reason why NC has inferred triggers from clean models is that there exist feature combinations that can achieve the evasion effect on clean models. 
Interestingly, the inferred trigger size is larger when NC uses the ``benign'' as the target class (which is also the real target class of the selective backdoor attack). This violates NC's expectation since NC assumes the trigger should be smaller for the truly poisoned class. Overall, the results confirm that our selective backdoor is stealthy against NC.

\section{MNTD Configurations}
\label{sec:mntd-config}
To adapt MNTD to work well on our dataset, we have communicated with the authors of MNTD. Based on the authors' suggestions, we configure MNTD as the following. We assume malware authors' goal is to let their malware samples evade the detection (instead of causing false positives). As such, we always set the target labels to ``benign'' for MNTD. Since our training samples are formatted as binary sparse feature vectors, we initialize the ``query set'' of MNTD accordingly, to mimic the feature distribution of the training set. Specifically, the query vectors are initialized by setting 10 to 100 random features to the value of 1, while the majority of the feature values are set to 0. This initialization method is used for both ``with query tuning'' and ``without query tuning'' settings. During meta classifier training, to achieve an effective AUC on the validation set, we also use a large query set of 100 inputs. Other parameters of MNTD follow the default setting of MNTD.

\begin{figure}[th]
    \centering
    \includegraphics[width=0.5 \linewidth]{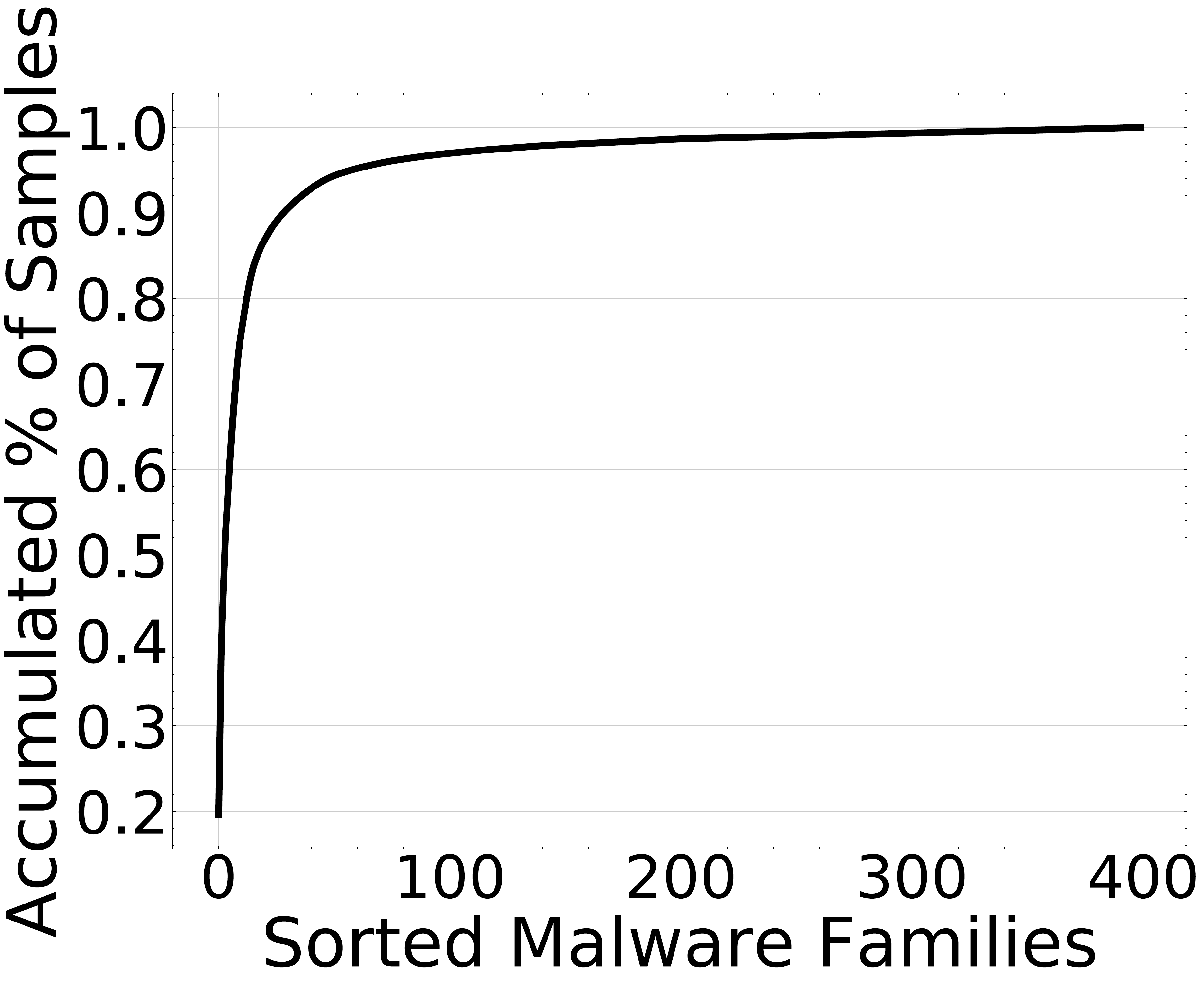}
    \caption{
    \textbf{Distribution of Malware Family Sizes}---%
    \textmd{{\small 
    we have 400 malware families in total. The top 13 families contribute 80\% of the malware samples.
    }}
    }
    \label{fig:cdf}
    \vspace{-0.13in}
\end{figure}

\section{Execution Time of JP Attack}
\label{sec:overhead}

In this section, we briefly discuss the computational overhead of the Jigsaw Puzzle (JP) attack. 

For the feature-space attack, the computational overhead primarily comes from Algorithm~\ref{alg:alter_optim} to optimize the trigger. For a given target family, the algorithm can converge within 2 hours. Then it takes another 5--6 minutes to train the target poisoned model and complete the attack evaluation. We run the feature-space experiment on a commodity server with Intel(R) Xeon(R) Silver 4214 CPU @ 2.20GHz, 192GB of RAM and Nvidia Quadro RTX 5000 GPU.

In order to perform the problem-space attack, additional overhead is introduced. First, we have a {\em preparation phase} that involves gadget harvesting, i.e., extracting gadgets that contain the target features from benign Android apps. For each feature, we consider a depth of 10 (i.e., searching 10 random benign apps). To complete the searching for all 10,000 features, it takes about 144 hours with a commodity server with 300GB of RAM and 48 cores Intel(R) Xeon(R) CPU E5-2697 v3 @ 2.60GHz. We argue that this is only {\em a one-time effort}---after the mapping between feature and bytecode gadget is created, they can be re-used to run future JP attacks for any target malware families.  

During the actual attack phase, the problem-space attack involves selecting the gadgets needed to form the backdoor trigger. Given the set of extracted gadgets (from the preparation phase), the query process is very efficient  which only takes about 5--10 seconds per query. 
This means that creating the problem-space trigger $\bM_{p}$ based on the feature-space trigger $\bM$ using Algorithm~\ref{alg:final_trigger} requires about at most 5 minutes for a trigger of size 30.

\end{document}